\documentclass[twocolumn,tighten]{aastex62}
\usepackage{textcomp}
\usepackage{gensymb}

\usepackage[hang,raggedright,tight]{subfigure}
\usepackage{longtable}
\usepackage[figuresright]{rotating}
\usepackage{float}
\usepackage[toc,page]{appendix}

\usepackage{gensymb}
\usepackage{threeparttable}
\usepackage{diagbox}
\usepackage{pbox}
\usepackage{booktabs}
\usepackage{multirow}  
\usepackage{makecell}
\usepackage{natbib}
\usepackage{amsmath}
\usepackage{CJKutf8}
\newcommand{\nxi}{\widehat{\xi}}

\shorttitle{AP test}
\shortauthors{Dong et al.}

\begin{document}
\title{
Tomographic
Alcock–Paczyński Test with Redshift-Space Correlation Function: Evidence for the Dark Energy Equation of State Parameter $w>-1$}
\author[0000-0003-0296-0841]{Fuyu Dong}
\altaffiliation{dongfy2020@kias.re.kr}
\affiliation{School of Physics, Korea Institute for Advanced Study (KIAS), 85 Hoegiro, Dongdaemun-gu, Seoul, 02455, Republic of Korea}
\author{Changbom Park}
\altaffiliation{cbp@kias.re.kr}
\affiliation{School of Physics, Korea Institute for Advanced Study (KIAS), 85 Hoegiro, Dongdaemun-gu, Seoul, 02455, Republic of Korea}
\author[0000-0003-4923-8485]{Sungwook E. Hong}
\affiliation{Korea Astronomy and Space Science Institute,
776 Daedeok-daero, Yuseong-gu, Daejeon 34055,
Republic of Korea}
\affiliation{Astronomy Campus, University of Science and Technology,
776 Daedeok-daero, Yuseong-gu, Daejeon 34055,
Republic of Korea}
\author[0000-0002-4391-2275]{Juhan Kim}
\affiliation{Center for Advanced Computation, Korea Institute for Advanced Study, 85 Heogiro, Dongdaemun-gu, Seoul, 02455, Republic of Korea}
\author{Ho Seong Hwang}
\affiliation{Astronomy Program, Department of Physics and Astronomy, Seoul National University, 1 Gwanak-ro, Gwanak-gu, Seoul 08826, Korea}
\affiliation{SNU Astronomy Research Center, Seoul National University, 1 Gwanak-ro, Gwanak-gu, Seoul 08826, Korea}
\author[0000-0002-7464-7857]{Hyunbae Park}
\affiliation{Lawrence Berkeley National Laboratory, CA 94720-8139, USA}
\affiliation{Berkeley Center for Cosmological Physics, UC Berkeley, CA 94720, USA}
\author{Stephen Appleby}
\affiliation{Asia Pacific Center for Theoretical Physics, Pohang, 37673, Korea}
\affiliation{Department of Physics, POSTECH, Pohang 37673, Korea}

\begin{abstract} 
The apparent shape of galaxy clustering depends on the adopted cosmology used to convert observed redshift to comoving distance, the $r(z)$ relation, as it changes the line elements along and across the line of sight differently. The Alcock-Paczyński (AP) test exploits this property to constrain the expansion history of the universe. We present an extensive review of past studies on the AP test. 
We adopt an extended AP test method introduced by Park et al. (2019), which uses the full shape of redshift-space two-point correlation function (CF) as the standard shape, and apply it to the SDSS DR7, BOSS, and eBOSS LRG samples covering the redshift range up to $z=0.8$.
We calibrate the test against the nonlinear cosmology-dependent systematic evolution of the CF shape using the Multiverse simulations. We focus on examining whether or not the flat $\Lambda$CDM `concordance' model is consistent with observation. 
We constrain the flat $w$CDM model to have $w=-0.892_{-0.050}^{+0.045}$ and $\Omega_m=0.282_{-0.023}^{+0.024}$
from our AP test alone, which is significantly tighter than the constraints from the BAO or SNe I$a$ methods by a factor of 3 - 6. When the AP test result is combined with the recent BAO and SNe I$a$ results, we obtain 
$w=-0.903_{-0.023}^{+0.023}$ and 
$\Omega_m=0.285_{-0.009}^{+0.014}$.
This puts a strong tension 
with the flat $\Lambda$CDM model 
with $w=-1$ at $4.2\sigma$ level. 
Consistency with $w=-1$ is obtained only when the Planck CMB observation is combined.
It remains to see if this tension between observations of galaxy distribution at low redshifts and CMB anisotropy at the decoupling epoch becomes greater in the future studies and leads us to a new paradigm of cosmology.

\end{abstract}
\section{Introduction}

\begin{table*}
\centering
\caption{Summary of the key AP test papers. PS, CF, and LSS stand for power spectrum, correlation function, and large-scale structures in the universe, respectively. $D_A$ is the angular diameter distance, and $H$ is the Hubble parameter.
  \label{Table-paper}}
\begin{minipage}{180mm}
\centering
\begin{tabular}{|c|p{4cm}|p{8cm}|c|}
\hline
\hline
Standard&Observables&Key references & \shortstack{Constrained\\quantity}\\
\hline
\multirow{7}{*}[-5ex]{\shortstack{Shape}} & Spherical objects & \cite{1979Natur.281..358A} &\multirow{7}{*}[-5ex]{$D_AH$}\\
\cline{2-3}
&Pairs of objects & {\cite{1994MNRAS.269.1077P,2010Natur.468..539M}}&\\
\cline{2-3}
&Cosmic voids & \cite{1995ApJ...452...25R,2012ApJ...754..109L}&\\
\cline{2-3}
&PS$^*$ of galaxies,  quasars & \cite{1996MNRAS.282..877B,2001MNRAS.328..174O}&\\
\cline{2-3}
&CF$^*$ of galaxies, quasars, Ly-$\alpha$ forests & \cite{1996ApJ...470L...1M,1999ApJ...518...24M,1999ApJ...511L...5H,2002MNRAS.332..311H} \cite{2012PhRvD..86f3503M}&\\
\cline{2-3}
&21-cm fluctuations & \cite{2005MNRAS.364..743N,2005MNRAS.363..251A,2006MNRAS.372..259B}&\\
\cline{2-3}
&Density gradient vector & \cite{2014ApJ...796..137L}&\\
\cline{2-3}
&Tomographic redshift-space CF$^*$& \cite{2015MNRAS.450..807L,2016ApJ...832..103L,2019ApJ...881..146P}&\\
\cline{2-3}
&Tomographic redshift-space PS$^*$& \cite{2019ApJ...887..125L}&\\
\hline
Population & Tomographic topology$^*$ of LSS & \cite{2010ApJ...715L.185P,2018ApJ...853...17A} & $D_A^2 /H$\\
\cline{2-3}
&Quasars& 
\cite{2020MNRAS.499L..36M}&\\
\hline
\end{tabular}
\begin{tablenotes}
      \small
      \item Note. * Measures of galaxy clustering shape are  used as observables in these works. 
\end{tablenotes}
\end{minipage}
\end{table*}

\cite{1979Natur.281..358A} studied the behaviour of the ratio $\Delta z /z\Delta \theta$ for spherical objects subtending $\Delta z$ along the line of sight and $z\Delta \theta$ across the line of sight. They found that the radial and angular separations depend on the cosmological parameters, such as the matter density parameter $\Omega_m$ and the cosmological constant $\Lambda$ differently, and the ratio is a sensitive function of the cosmological constant. The finding led them to propose to use this ratio to test for non-zero cosmological constant. Being a geometry measure, the ratio is supposed to be independent of various astrophysical evolutionary effects even though statistical separation of the peculiar velocities from the Hubble expansion remained a major obstacle for application of the test.

This idea (hereafter `the AP test')  is now more than 40 years old. Since the test has been suggested, many methods and observables have been proposed for practical applications. As a result, the test has been getting more powerful, and is producing interesting constraints on cosmological parameters in recent years. Unlike other cosmological probes used to constrain the expansion history of the universe, however, in the case of the AP test many different kinds of cosmic objects are being used and the statistical measures adopted are diverse. Therefore, it is worthwhile to review the literature on the AP test at this point in order to learn about the past development of this test and to find the desirable directions of the future studies. In the next section we will review the literature in roughly chronological order and list mainly the first papers introducing new method or tracers of the AP effect in Table 1. A point to pay attention to is how the cosmological geometric distortion has been separated from the distortion due to the peculiar velocity. 

In this paper, we will carry out the AP test using one of the standard shapes\footnote{The name `standard shape' is used here as the shape of cosmic objects that is conserved in time and can be used to examine through the AP test if the adopted cosmology is correct or not. The name intends to extend the choice of the particular shape, i.e. spherical shape, which has been adopted as the shape of the standard objects of the AP test in previous works.}, the full shape of the redshift-space two-point correlation function (CF) along and across the line of sight. This tomographic method has been proposed and developed by \cite{2010ApJ...715L.185P}, \cite{2014ApJ...796..137L,2015MNRAS.450..807L,2016ApJ...832..103L,2018ApJ...856...88L}, and \cite{2019ApJ...881..146P}. The SDSS galaxy redshift surveys such as DR7 Main galaxies (KIAS-VAGC, \cite{2010JKAS...43..191C}), BOSS galaxies \citep{2015ApJS..219...12A}, and eBOSS Luminous Red Galaxies (LRGs, \cite{2016ApJS..224...34P}) are used in the test. We constrain the flat $w$CDM model by looking for the cosmological parameters that result in non-evolving shape of the redshift-space CF. The main focus of this paper is to examine whether or not the current standard cosmological model, the flat $\Lambda$CDM model, is consistent with observation. Mock survey samples derived from a set of high-resolution large-volume cosmological $N$-body simulations are used to accurately mimic the observational sample selection (thus galaxy biasing), non-linear gravitational effects, and also cosmology dependence of their redshift evolution.

This paper is organized as follows. 
Section 3 introduces the survey data used for our analysis. \S\ref{sec:simudata} introduces the simulation data, and \S\ref{sec:method} describes the methodology.  \S\ref{sec:result} shows our results for constraining $w$CDM model. We discuss and conclude in  \S\ref{sec:conclusion}.

\section{A review of the AP test-related studies}

Up to now various cosmological `objects' have been suggested to use for the AP test. Table \ref{Table-paper} summarizes most of the standard shapes and populations suggested for the AP test and volume test associated with the AP distortion. References are some of the key papers that first proposed the ideas.

Alcock \& Paczyński's proposal for using geometrical probe to constrain the cosmological constant was ignored for a long time after their pioneering paper was published \footnote{In October 1990, Bohdan Paczyński invited one of the authors (CBP) and handed a copy of his AP paper. He strongly recommended to work on this test. He said the paper had an excellent idea but nobody was working on it. Our work has been motivated by his recommendation, 
and this paper is devoted to Bohdan.}, and it took 15 years for the community to finally pay attention to this idea.  

\cite{1994MNRAS.269.1077P} has suggested to use physical pairs of quasars and inspected the angle between the line of sight and the separation vector of pairs. If quasar pairs are randomly oriented, the angles will have a uniform distribution when the correct cosmology is adopted. However, the impact of peculiar velocities of quasars and their systematic redshift evolution were not studied.
 
\cite{1995ApJ...452...25R} has chosen cosmic voids, and used the volume and the axis ratio of voids to constrain the deceleration parameter $q_0$. It was pointed out that the test is limited by the intrinsic scatter in the size and shape of the voids and the peculiar velocities will make additional contribution to the uncertainty.
 
\cite{1996ApJ...470L...1M} and \cite{1996MNRAS.282..877B} have proposed to use the clustering of galaxies or quasars assuming its statistical isotropy, and have taken into account the linear redshift-space distortion effect of peculiar velocities of the tracers. Matsubara \& Suto (see also \cite{1998ApJ...494...13N} and \cite{1998ApJ...498...11P}) proposed to make simultaneous measurement of redshift-space distortion in the CF of galaxies at low redshifts (dependent mostly on $\Omega_m$) and of quasars at high redshifts (depends on both $\Omega_m$ and $\Lambda$) to constrain these cosmological parameters. It was shown that $\Lambda$ can be measured by disentangling the cosmological shape distortion from the distortion due to peculiar motion of galaxies (redshift-space distortion, hereafter RSD). Ballinger et al. chose to use the isotropy of the power spectrum instead of CF because the modelling of RSD is simpler in Fourier space than in real space. The correct cosmological model can be searched for as adopting a wrong geometry, i.e. wrong $r(z)$ relation, makes a spherical object appear distorted. Both linear RSD due to peculiar velocities induced by gravitational instability \citep{1987MNRAS.227....1K} and the nonlinear distortion due to Fingers-of-God caused by orbital motions of galaxies in galaxy groups and clusters \citep{1992MNRAS.259..494P,1994ApJ...431..569P} are taken into account.
 
\cite{1999ApJ...518...24M} has suggested to disentangle the effects of geometry and peculiar velocities using the CF of the Ly$\alpha$ forest clouds along close pairs of quasars. It is assumed that the Ly-$\alpha$ forests closely represent the underlying matter density distribution. And the linear theory for the redshift-space CF of the matter density field is used to model the RSD effect. They also pointed out that the constraint on the parameter $c^{-1}H(z)D_{A}(z)$ can be much tighter if the parameter $\beta=f/b$ can be estimated independently, where $f$ is the growth rate function, and $b$ is the galaxy bias factor. \cite{1999ApJ...511L...5H} has also chosen the Ly-$\alpha$ forest clouds toward close quasar pairs to measure CFs along and across the line of sight to obtain the auto-spectrum and cross-spectrum. By assuming a cosmological model, the auto-spectrum can be converted to three dimensional power spectrum in an integral form with a kernel that depends on the peculiar motion and thermal broadening. The adopted cosmological model is then constrained through comparisons between the predicted and observed cross-spectrum.

\cite{2000ApJ...528...30M} has inspected the effects of nonlinear density and velocity fields on the cosmological redshift-space distortion in the power spectrum. The nonlinear effects are estimated by using $N$-body simulations of SCDM, LCDM, and OCDM models, and compared with theoretical predictions.

\cite{2001MNRAS.328..174O} has applied the method of Ballinger et al. using the distortions in the PS parallel and perpendicular to the line of sight, $P^s (k_{\parallel}, k_{\perp})$, to the 2dF QSO redshift survey data, and obtained a joint-constraint in the $\Omega_m$-$\beta$ space yielding $\beta = 0.39^{+0.18}_{-0.17}$ and $\Omega_m (z=0) = 0.23^{+0.44}_{-0.13}$ for a flat $\Lambda$CDM model. They later revised the results using the final 2dF QSO catalog to obtain $\Omega_{\Lambda}=0.71^{+0.09}_{-0.17}$ for the flat $\Lambda$ cosmology \citep{2004MNRAS.348..745O} favoring non-zero $\Lambda$. \cite{2002MNRAS.332..311H} used the same 2dF QSO data and calculated the CFs parallel and perpendicular to the line of sight. An iterative procedure was used to find the best fit cosmology and $\beta ({\bar z})$ at the fixed mean QSO redshift, and resulted in a solution marginally favoring non-zero $\Lambda$ model.

\cite{2005MNRAS.364..743N} examined the possibility of applying the AP test on the three-dimensional map of 21-cm emission at the epoch of reionization, and concluded that at $z\sim 20$ the ratio of the redshifted 21-cm emission frequency intervals to angular separations is sensitive to the assumed cosmological parameters and the dark energy equation of state. \cite{2005MNRAS.363..251A} also suggested to use the CF of the HI distribution during reionization and post-reionization epochs. The total hydrogen content is assumed to trace the dark matter, and the reionization proceeds through spherical bubbles of ionized gas. The HI gas in the post reionization era is assumed to be in high column density clouds, which could be biased with respect to the underlying dark matter. \cite{2006MNRAS.372..259B} chose the 21-cm power spectrum for separation of the AP anisotropy from other effects and a separate measurement of the Hubble parameter. This method was revisited by \cite{2020MNRAS.499..587E} using a hydrodynamical cosmological simulation to obtain the 21cm signal map.

\cite{10.1111/j.1365-2966.2006.11168.x} investigated the AP test using galaxy clusters. It is shown that, if the peculiar velocities of galaxy clusters can be measured using the kinetic Sunyaev-Zel'dovich effect, the galaxy cluster CF becomes a pure test of cosmic geometry. 

A different application of the AP test has been suggested by \cite{2010ApJ...715L.185P} and \cite{2014ApJ...796..137L,2015MNRAS.450..807L,2016ApJ...832..103L},
where the full redshift dependence of the standard shape, rather than the shapes at individual epochs,
is used to constrain the expansion history of the universe. \cite{2010ApJ...715L.185P} proposed to use the topology of large-scale structures in the universe as the `standard population' because it is insensitive to the nonlinear gravitational evolution, galaxy bias, and RSD.  Because the genus is a local quantity calculated from integration of local curvature of iso-density contour surfaces, it can be independent of survey boundary effects. Non-uniformity of angular or radial selection function can be easily incorporated in the analysis.  It is noted that, when a wrong cosmology is adopted in the conversion from redshift to comoving distance, the genus, being the count of intrinsic topological objects (i.e. large-scale structures in the universe), is erroneously altered due to the change of local volume and smoothing scale. The standard population is required to be independent of redshift on the observed past light cone surface \citep{1992ApJ...387....1P}. The correct cosmology is found through an iterative procedure looking for the case resulting in no redshift evolution of topology. 
This method of using tomographic measurement of topology of a standard population has been applied to the AP test by \cite{2014ApJ...796..137L} using the density gradient vector as the observable, and by \cite{2015MNRAS.450..807L,2016ApJ...832..103L} who adopted the redshift-space galaxy CF.

\cite{2010PhRvD..81d3512S} pointed out that it is difficult to distinguish between dark energy and modified gravity using RSD because peculiar motions and the AP effect jointly affect the RSD, and modified gravity can be degenerate with the unknown cosmological parameters.

\cite{2010Natur.468..539M} used gravitationally bound pairs of galaxies in the SDSS ($z < 0.05$) and in the DEEP2 survey ($z\lesssim 1.45$) to constrain cosmological parameters and found that the flat universe is preferred ($\Omega_k = 0.03\pm 0.12$). In the case of the flat universe they obtained $0.6 < \Omega_X <0.8$ and $-1.12 < w_X < -0.85$, where $X$ stands for dark energy.

\cite{2011MNRAS.418.1725B} combined the AP test applied to galaxy power spectra measured from the WiggleZ Dark Energy Survey data ($0.1<z<0.9$) with supernova observations to find that ${\ddot a}$ has been positive (i.e. the universe has been accelerating) and, in a flat $\Lambda$CDM model, $\Omega_m = 0.29 \pm0.03$. Linear galaxy biasing is assumed and an empirical model is adopted for large-scale RSD. For the small-scale RSD the Lorentzian damping function is chosen (Park et al. 1994).

\cite{2012MNRAS.420.1079J} proposed a way to improve the AP test using the angles subtended by galaxy pairs. They pointed out that the comoving separation of pairs and their radial peculiar velocities can be correlated, and suggested to numerically measure the dependence of the normalization parameter, used to transform the true angle distribution into the apparent one, on redshift and cosmology, which has been neglected by \cite{2010Natur.468..539M}

\cite{2012ApJ...754..109L} developed a prescription on using the shape of stacked voids in a few redshift shells for the AP test. It is assumed that the void statistics from $N$-body simulations is preserved for the voids in the galaxy distribution. The impact of peculiar velocities on distortions is measured from simulations, for example. \cite{2012ApJ...761..187S} has applied this method to the stacked voids in the SDSS main and luminous red galaxies searched for out to $z=0.36$. However, the stretch factor was too noisy to constrain cosmology. \cite{2014MNRAS.443.2983S} used an extended watershed algorithm to identify about 1500 cosmic voids out to $z\sim 0.6$ in SDSS DR7 and DR10 LOWZ/CMASS catalogs. They claimed that stacked voids showed a universal flattening of 14\% along the line of sight due to peculiar velocities for all void sizes at all redshifts and that the AP signal was detected in the sample with a high statistical significance. This method still suffered from insufficient statistics and accurate modeling of peculiar velocity effects. The same void finding algorithm was used by \cite{2017ApJ...835..160M}, who used the SDSS DR12 BOSS to identify voids at $z=0.43-0.7$ and obtained a constraint $\Omega_m=0.38_{-0.15}^{+0.18}$ using mock catalogs assuming flat $\Lambda$CDM model. It was found that the `shape noise' due to variation of individual void shapes makes the void method significantly weaker than predicted.

\cite{2012PhRvD..86f3503M} proposed a new method of carrying out the AP test without adopting trial cosmologies, but using the RSD in clustering statistics in angular and redshift separations.

\cite{2014ApJ...781...96L} has used the galaxies in SDSS BOSS DR10 catalog and the quasars in SDSS DR7 and BOSS catalogs to calculate the correlation function in $(\Delta z, z\Delta \theta)$ space to discriminate between different cosmological models. Adopting the flat $\Lambda$CDM model, they obtained $\Omega_m = 0.24_{-0.07}^{+0.10}$.


\cite{2014ApJ...796..137L} proposed to use the galaxy mass density gradient field in redshift space as the standard shape for the AP test. The anisotropy in the density gradient field at a given redshift is produced by both peculiar velocities and the AP effects. Li et al., following the idea of \cite{2010ApJ...715L.185P}, used the redshift dependence of this anisotropy, which is dominated by the geometrical AP effect. The degeneracy direction of the likelihood contour from this standard shape is nearly orthogonal to that from the correlation function method of \cite{2015MNRAS.450..807L,2016ApJ...832..103L}.

\cite{2015MNRAS.450..807L} chose the angular shape of the redshift-space CF, $\xi(\mu)$, as the standard shape, where $\mu={\rm cos} \theta$ and $\theta$ is the angle between galaxy separation and the line of sight. 
The redshift-space CF
was found to exhibit only a small amount of redshift evolution compared to the AP effect even though it is significantly distorted due to the peculiar motion of galaxies. It was also suggested to use the volume effect on $\xi(\mu)$ due to the AP effect to additionally constrain cosmological parameters.  \cite{2016ApJ...832..103L} applied this method to SDSS-III BOSS DR12 LOWZ and CMASS samples and obtained very tight constraints on the cosmological parameters governing the expansion history of the universe: $\Omega_m=0.301\pm 0.006, w=-1.054 \pm0.025$ for the flat $w$CDM universe when constraints from other probes (Planck CMB in particular) are combined together. All the nonlinear effects on the redshift evolution of $\xi(\mu)$ were estimated from mock survey samples extracted from large cosmological simulations that modeled the SDSS samples. And the part of the CF with the separation $6 h^{-1}$Mpc $<s< 40  h^{-1}$Mpc was used in the test, which provided much higher statistics compared to the methods using the BAO.

\cite{2017JCAP...02..020L} studied the AP test applied to the BAO feature in the CF by using a formalism that included relativistic effects. It was found that the gravitational lensing correction to the galaxy number density fluctuation affects only the amplitude of the CF around the BAO scale but not its shape.

\cite{2018ApJ...856...88L} extended their cosmological AP test to the dynamical dark energy models with $w=w_0+w_a z/(1+z)$. Using the redshift dependence of the redshift-space CF $\xi(\mu)$ measured from the SDSS DR12 BOSS sample, they found $\Omega_m=0.301\pm0.008, w_0 =-1.042\pm0.067$, and $w_a=-0.07\pm 0.29$ when combined with other cosmological probes. It was also found that adding the AP test to the SN Ia, CMB, and BAO results reduces the uncertainties in cosmological parameters by 30\%–40\% and improves the dark-energy figure of merit by a factor of about 2. The test was further extended to other cosmological parameters by Li et al. (2018b) and \cite{2019MNRAS.483.1655Z}, who reported $H_0 = 67.8_{-1.9}^{+1.2}$ km s$^{-1}$ Mpc$^{-1}$ by combining constraints from the tomographic CF AP method and BAO measurements. To make estimation of the systematics in the AP distortion feasible for a large class of cosmological models \cite{2020ApJ...890...92M} studied the redshift evolution of the RSD and the effect of halo bias on $\xi (\mu, z)$ using a fast $N$-body simulation algorithm that combines the second-order Lagrangian perturbation theory on large scales and the Particle-Mesh $N$-body code on small scales.

\cite{2019ApJ...881..146P} improved the method of \cite{2015MNRAS.450..807L,2016ApJ...832..103L} and presented a method using the full two-dimensional shape information of the normalized redshift-space CF. By using a set of $N$-body simulations of different cosmological models (a library of the `Multiverse Simulations') cosmology dependence of the nonlinear evolution of the CF shape was measured.  They found that using the information on ${\hat \xi}(r, \mu, z)$ in the tomographic AP test tightens the constraints on $\Omega_m$ and $w$ by about 40\% compared to the method using ${\hat\xi}(\mu, z)$.

\cite{2019ApJ...887..125L} studied the feasibility of using the redshift evolution of the galaxy power spectrum ${\hat P}_{\Delta k}(\mu)$ for conducting the tomographic AP test. They found that the impacts of halo bias and RSD play only minor role in the shape distortion in the Fourier space compared to the AP effect. 

\cite{2019ApJ...878..137Z} tried to reconstruct the dark energy equation-of-state $w$ in a non-parametric form and favored a dynamical dark energy at $z<1$ when the AP method is applied to recent observational data sets.

\cite{2020MNRAS.499L..36M} used a count-in-cell method measuring the height of two adjacent cells along the line of sight and applied the method to the SDSS-IV eBOSS quasars to favor the $\Lambda$CDM model against others. It was assumed that the quasar number density does not evolve across neighboring cells.

Recently, \cite{2021PhRvD.103l3534S} introduced tripolar spherical harmonic expansion of the CF as a way to capture its angular dependence due to the wide-angle AP effect. They reported that the error in the density CF exceeds 10\% when the opening angle $\theta > 30^\circ$.

In this paper we measure the redshift evolution of the shape of CF using the galaxies in SDSS DR7, BOSS LOWZ and CMASS, and eBOSS LRG samples.
The small intrinsic systematic evolution of the CF shape due to cosmology-dependent nonlinear gravitational evolution and galaxy bias variation across observational samples is accurately estimated by using the Multiverse simulations. Applying \cite{2019ApJ...881..146P}'s method of using both angular and radial shape of the galaxy CF, we constrain the flat $w$CDM model to have $w=-0.892_{-0.050}^{+0.045}$ and $\Omega_m=0.282_{-0.023}^{+0.024}$
using the AP test alone.

\section{Observational Data}
\label{sec:data}
\begin{figure*}[!htb]
    \centering
     \subfigure{
     \includegraphics[width=1\linewidth, clip]{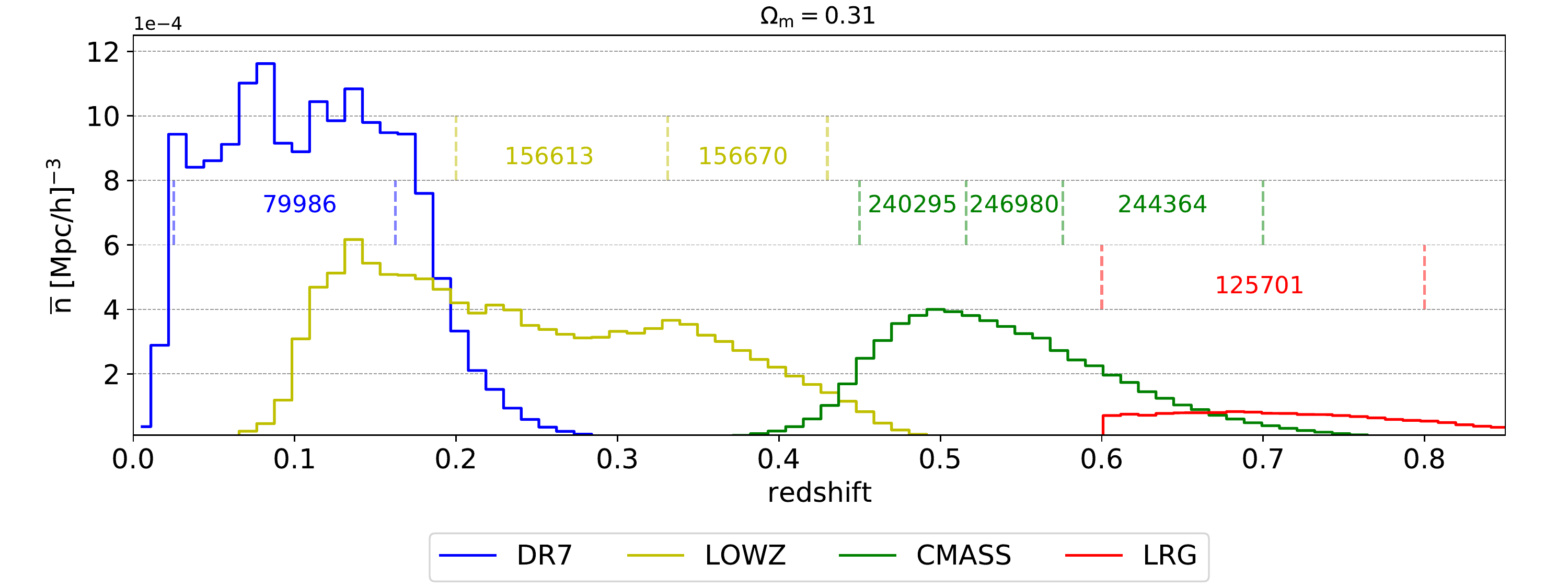}}
   \caption{
   Distribution of galaxy number density in the SDSS samples used in this study. A $\Lambda$CDM cosmology with $\Omega_m=0.31$ is used. The blue, yellow, green and red color histograms are for the DR7, LOWZ, CMASS, and eBOSS LRG samples, respectively. Vertical dashed lines mark the borders separating seven redshift slices. The same color is used as for each parent galaxy sample. 
   We select the DR7 galaxies with $M_r < -21.07$. 
   The LOWZ galaxies with $M_*>10^{11.1}M_\odot$ are used while the CMASS and eBOSS LRG galaxies are selected to have $M_*>10^{11}M_\odot$. The number of galaxies contained in each redshift slice is given.
   It can be seen that the eBOSS LRG sample has galaxy number density significantly lower than the CMASS sample.  }
    \label{fig:ngal}
\end{figure*}

We use a set of spectroscopic samples of the SDSS surveys in our analysis. 
The lowest redshift sample is the SDSS-I/II DR7 Main galaxy sample, and we make use of the Korea Institute for Advanced Study Value-Added Galaxy Catalog (KIAS-VAGC, \cite{2010JKAS...43..191C}). 
This catalog is based on the New York University Value-Added Galaxy Catalog \citep{2005AJ....129.2562B} as part of the SDSS DR7 (\cite{2009ApJS..182..543A} but supplements missing redshifts taken from various galaxy redshift catalogs for a higher redshift completeness. The sky coverage of KIAS- VAGC is about 8000 ${\rm deg^2}$.

For the deeper SDSS-III BOSS sample, we use the catalog generation of the Data Release 12 \citep{2015ApJS..219...12A}, for which spectra and redshifts of galaxies are obtained over 9400 ${\rm deg^2}$.
The galaxy number density of the BOSS sample is typically $3\times 10^{-4}(h^{-1}{\rm Mpc})^{-3}$, which is useful for large-scale cosmological studies \citep{2016MNRAS.455.1553R}. It consists of two subsamples: `LOWZ' ($0.1\lesssim z\lesssim0.4$) and `CMASS' ($0.4\lesssim z\lesssim0.7$). The LOWZ sample is an extension of the fainter SDSS I/II LRG sample. The CMASS sample is intended to select a stellar mass-limited sample of galaxies.

Our highest redshift sample is the Data Release 16 of the eBOSS LRG sample. The goal of the LRG selection is to obtain a sample at redshifts greater than the BOSS CMASS sample. There are two types of LRG catalog released: the ``full" data catalog containing all the information on targets that pass the veto masks and the ``clustering" data catalog containing only those with good redshifts after mask, completeness, and redshift cuts are all applied. In this work, we adopt the eBOSS LRG ``clustering" catalogue (eLRGc afterwards) for the CF statistic, which occupies $\sim 4000 \,{\rm deg}^2$ on the sky. In Appendix \ref{sec:lrg-angular} we build the angular mask of the eLRGc sample, which is used to create mock samples in the next section.

Figure \ref{fig:ngal} shows the redshift distribution of galaxy number density for the samples used in our analysis. It can be noted that the number density of eLRGc is much lower compared to those of other samples. As each sample was obtained by targeting galaxies using different criteria and algorithms, we perform our analysis separately for each sample. 
For DR7 galaxies \footnote{The absolute magnitude of $M_r=-21.07+5 {\rm log} h$ corresponds to $M_*\sim 10^{11} M_\odot$ for the DR7 sample. 
}, we apply a $r$-band absolute magnitude cut of
$M_r<-21.07+{\rm 5log} h$
to construct a volume-limited sample spanning the redshift range of $0.025<z<0.163$. For the remaining deeper samples we put a lower stellar mass cut. This is to use only the galaxies within the range with relatively high stellar mass sampling rate. The prescription also removes the galaxies that suffer from large stellar mass uncertainties.
Specifically, for the LOWZ sample, we apply the low mass cut of $M_*>10^{11.1} M_\odot$. For the CMASS and LRG samples, a slightly lower cut of $M_*>10^{11} M_\odot$ is applied, considering their lower mass distributions compared to LOWZ.
We split the galaxies in the four catalogs into seven redshift bins: one bin in DR7 ($0.025<z_1<0.163$), two bins in LOWZ ($0.2<z_2<0.331<z_3<0.43$), three bins in CMASS ($0.45<z_4<0.516<z_5<0.576<z_6<0.7$), and one in eLRGc ($0.6<z_7<0.8$). We consider the lower six redshift slices as our baseline galaxy samples, regarding their much higher number densities than LRGs.

Stellar mass of galaxies are obtained from various sources. For the DR7 sample, we adopt the galaxy stellar mass derived in the MPA-JHU catalogue \footnote{https://wwwmpa.mpa-garching.mpg.de/SDSS/DR7/} as a reference. Following \cite{2018ApJ...858...30G}, we reduce the galaxy stellar mass of \cite{10.1111/j.1365-2966.2011.20306.x} by $\sim$0.155 dex to be consistent with the literature\footnote{The stellar mass with the PCA method is systematically overestimated by 0.105 dex compared to the MPA-JHU stellar mass and a further constant shift of -0.05 dex is considered when converting the IMF from a \cite{2001MNRAS.322..231K} to that of Chabrier(2003).}. 
The stellar mass of the BOSS galaxies used here was estimated by \cite{10.1111/j.1365-2966.2011.20306.x}, where the galaxy spectra are fit by using the principal component analysis (PCA). We use the stellar mass obtained by applying the stellar population synthesis (SPS) models of the \cite{2003MNRAS.344.1000B} SPS model  with the IMF of \cite{2001MNRAS.322..231K} and the dust attenuation model of \cite{2000ApJ...539..718C}.
We estimate stellar mass of the LRG galaxies with the CIGALE routine \citep{2019A&A...622A.103B}, for which we refer readers to Appendix \ref{sec:lrg-ms} for more details.

The random catalogs corresponding to the LOWZ, CMASS and eLRGc samples with the angular selection functions incorporated are provided along with the data. We extract the random points from these catalogs according to the redshift distribution of the galaxy samples used here.
Each of the BOSS target and eBOSS target are assigned with three types of weight: `WEIGHT\_CP', `WEIGHT\_NOZ' and `WEIGHT\_SYSTOT', aiming at removing the systematics in the data introduced by close pairs, redshift inefficiency as well as the imaging systematics. In our analysis, the total weight for each galaxy/random point is the combination of the three weights: $(w_{\rm cp}+w_{\rm noz}-1)w_{\rm sys}$ for BOSS target and $w_{\rm cp}w_{\rm noz}w_{\rm sys}$ for eBOSS target \footnote{eBOSS uses a different weighting scheme from BOSS. For example, BOSS simply upweight the `nearest-neighbour' (NN) galaxy from the same target class that was assigned a fibre to account for collided galaxies that were not assigned fibres\citep{2016MNRAS.455.1553R}, which is tracked by a weight $w_{\rm cp}$. The CP correction in eBOSS is a variant of the standard NN method \citep{2020MNRAS.498.2354R}. In particular, $w_{\rm cp}$ weights are computed for collision groups where the weight of each target missed due to a fibre collision is equally distributed among the observed members of the group
rather than assigning it only to its nearest neighbour as in the standard NN method.}.

The SDSS spectroscopic galaxy surveys suffers from the fiber collision problem as some  target galaxies are not assigned fibers due to the mechanical limitation of the spectroscopic instrument.
The fiber collision effect can affect measurements of small-scale clustering of galaxies. Such systematics should be corrected for an accurate shape measurement of CF. We employ the mock survey data explained in the next section to model the effects and estimate the systematics accurately. The details are given in Appendix \ref{sec:fiber}.

\section{Simulation Data}
\label{sec:simudata}
\begin{figure}[!htb]
    \centering
     \subfigure{
     \includegraphics[width=0.85\linewidth, clip]{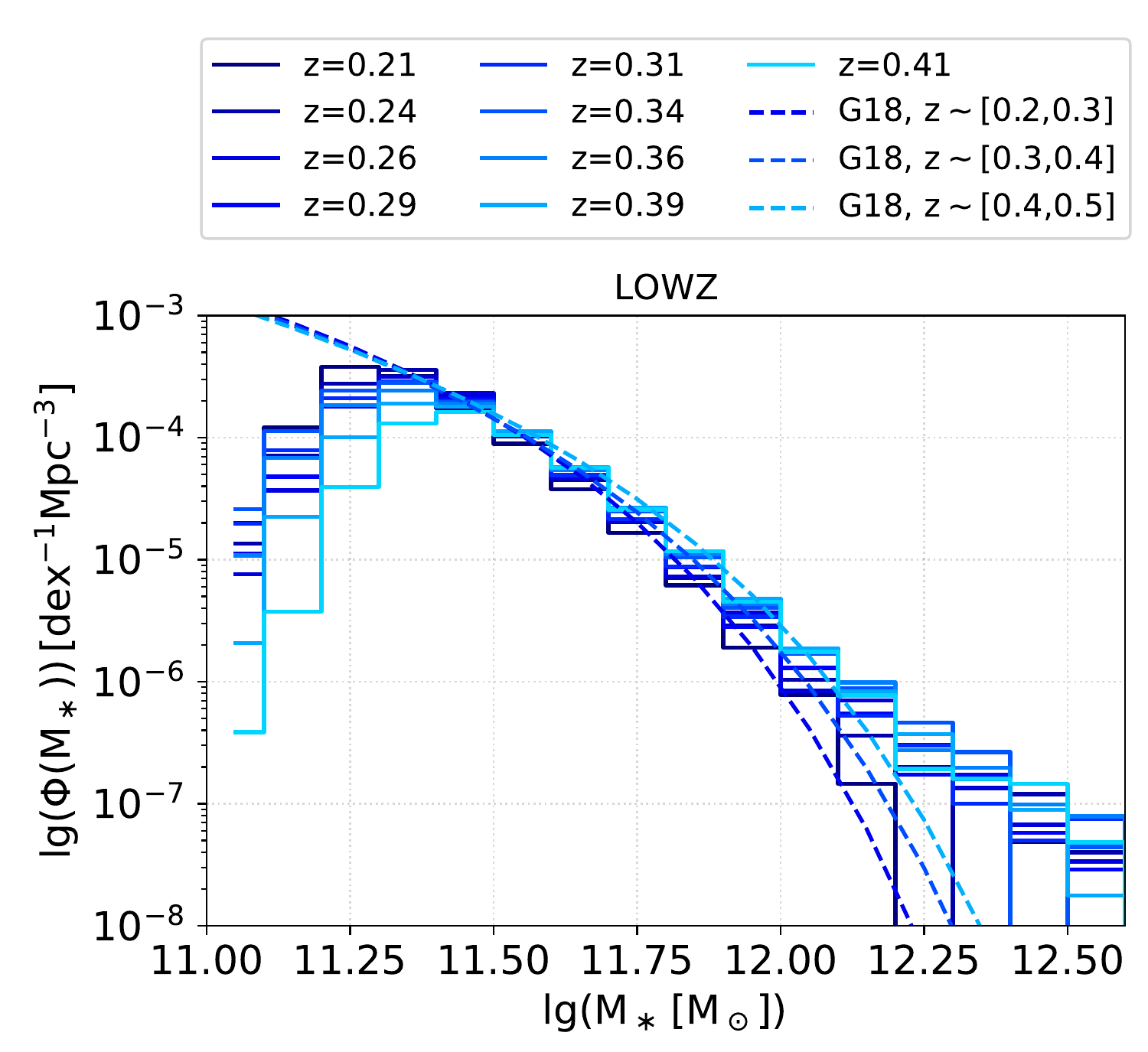}}
     \subfigure{
     \includegraphics[width=0.85\linewidth, clip]{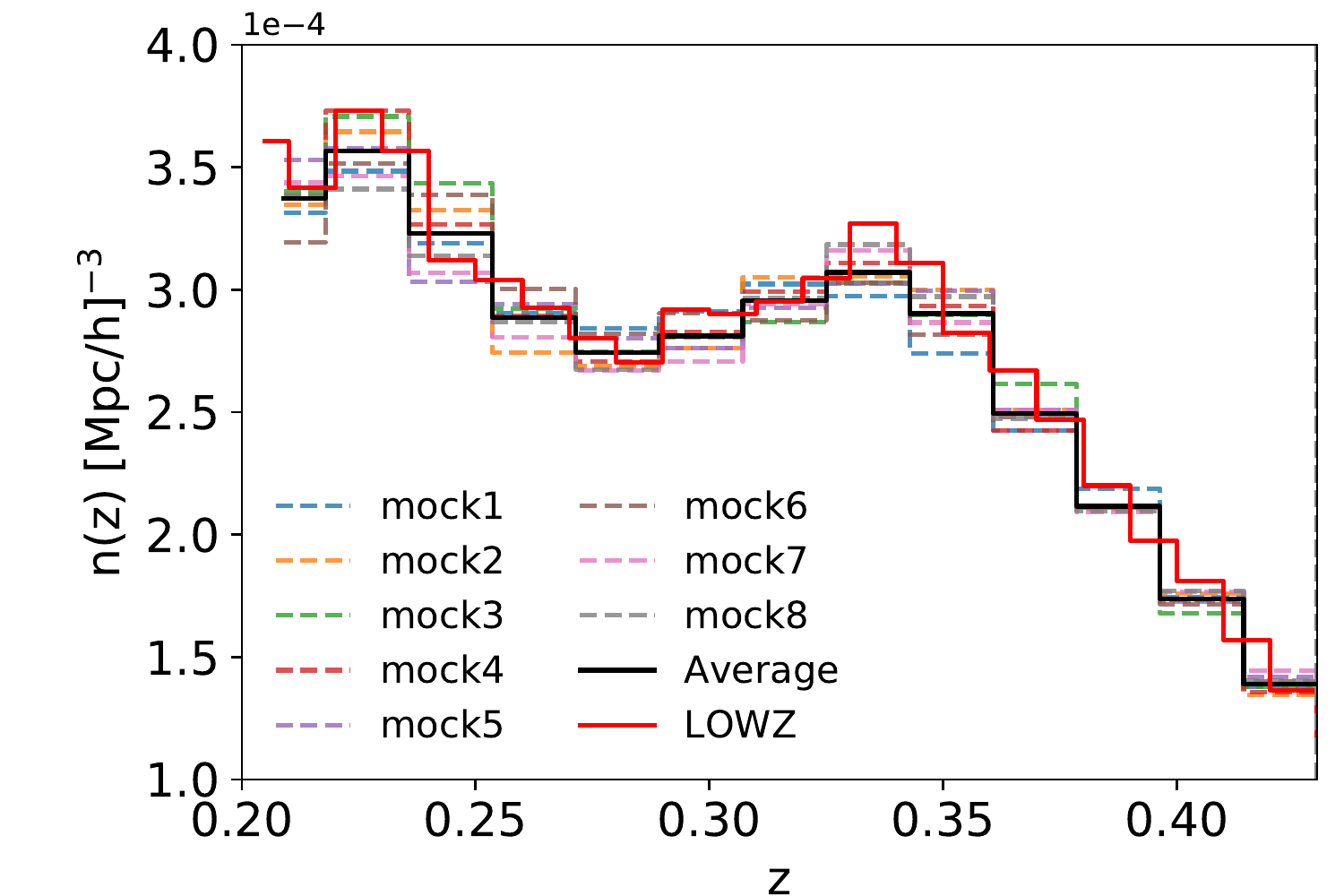}}
   \caption{Upper panel: stellar mass function of the LOWZ sample. The solid line presents the observed stellar mass functions of the LOWZ sample in various redshift bins. Mass selection is clearly incomplete at the low-mass end, which is more pronounced at higher redshifts. We adopt the complete stellar mass function reconstructed by G18 (dashed lines for three redshift intervals) for performing the subhalo abundance-matching in simulation. We can obtain the mass-selection function at each redshift bin by comparing the observed function to the dashed line.  Lower panel: galaxy number density distribution of the LOWZ sample in redshift (red solid line). The dashed lines are those for eight HR4 mocks by adopting the mass-selection function. The black solid line shows the mean of all mocks, which well restores the observational data.}
    \label{fig:SMF}
\end{figure}

In addition to the AP effect, the peculiar motion of galaxies can produce redshift-dependent anisotropic clustering. 
This is the main source of systematic uncertainties in our method as in all the AP effect-based tests.
The observed redshift of a galaxy, $z_{\rm obs}$, is a combination of the cosmological contribution, $z_{\rm cos}$, arising from the Hubble expansion with the Doppler effect, $z_{\rm pec}$, induced by the line of sight component of the galaxy’s proper peculiar velocity, $v_{\rm pec}$:
\begin{equation}
\label{eq:zrsd}
    z_{\rm obs}=z_{\rm cos}+\frac{v_{\rm pec}}{c}
    (1+z_{\rm cos}),
\end{equation}
which shows the impact of peculiar velocity on redshift-space position of galaxies. The evolution of pairwise peculiar velocity of galaxies can introduce evolution in the shape of redshit-space CF in both linear and non-linear scales. This change of shape is much smaller than the linear RSD itself, but is still important and must be corrected for precision measurement of cosmological parameters.
In addition, the angular and radial survey selection functions of galaxy survey samples with different degree of galaxy bias could also introduce redshift-dependent systematics to the CF. We will employ two sets of simulations to accurately model and estimate these systematics: the Horizon Run 4 (HR4) and Multiverse simulations.

\subsection{Horizon Run 4}
\label{sec:hr4}
The HR4 simulation has a box size of 3150 $h^{-1}{\rm Mpc}$, and employs $N=6300^3$ dark matter particles \citep{2015JKAS...48..213K}. It adopts the WMAP 5-year cosmology, with $(\Omega_{\Lambda},\Omega_m,\Omega_b,h,\sigma_8,n_s)=(0.74,0.26,0.044,0.72$, $0.794,0.96)$. We take the most-bound dark matter particle of each halo as the galaxy \citep{2016ApJ...823..103H}, and produce mock `galaxies' by using the modified merger time scale model of \cite{Jiang_2008},
\begin{equation}
    \frac{t_{\rm merge}}{t_{\rm dyn}}=\frac{(0.94\epsilon^{0.6}+0.6)/0.86}{{\rm ln}[1+(M_{\rm host}/M_{\rm sat})]} \left(\frac{M_{\rm host}}{M_{\rm sat}}\right)^{1.5}.
\end{equation}

To construct realistic mock catalogs we convert the real-space positions of galaxies in the simulation into redshift-space positions in a spherical thick shell that conforms to the geometry of each survey. In this process, galaxies are drawn from the output snapshot of the simulation that is closest in epoch to the observed sample\footnote{Snapshot data for HR4 simulation is available at z=0, 0.05, 0.1, 0.15, 0.2, 0.3, 0.4, 0.5, 0.6, 0.7, and 1}. Taking the LOWZ sample as an example, we choose the snapshot with redshift $z=0.3$ to produce the mock samples. 
We first put the bottom half of the simulation box along the line-of-sight (LOS) on one face. We then rotate the box to match the orientation of mock galaxies to the LOWZ sample. Next, we convert the LOS galaxy distance into a redshift using the distance–redshift relation corresponding to the simulation cosmology. The peculiar velocity of the most-bound dark matter particle is assigned to the corresponding galaxy. We add the peculiar velocity contribution to the redshift of each object according to Equation \ref{eq:zrsd} to mimic the RSD effect, and apply the survey angular selection function. 
The second mock sample is made 
by translating the observer position across the top half of the cube box in the survey direction.  To make more mocks, we gradually move the observer position within the simulation box along the LOS. Each time we change the observer location, we 
ensure that the spherical shell does not re-sample the volume taken by previous mocks. For higher redshift samples, such as the eBOSS LRG, we stack the simulation box in the direction perpendicular to the face to cover the survey area that increases with distance. Through this procedure, we create 8 mocks for BOSS samples and 6 mocks for the eLRGc samples. In this work we use these mock samples to estimate the systematics in the observed CF.

To match the number density of `mock' galaxies to that of BOSS and eBOSS galaxies, we use stellar mass as the total mass indicator. The observed stellar mass function $\phi(M_*,z)$ of the LOWZ galaxies is shown in the upper panel of Figure 2 in each thin redshift bin with the width of $\Delta z=0.03$. We assign stellar mass to each mock galaxy according to these stellar mass functions, using the subhalo abundance-matching technique \citep{2004MNRAS.353..189V}. 
More specifically, the connection between  mock and SDSS galaxies are built up by comparing the number density of mock galaxies at $z$ with mass greater than $M$ to the number density of observed galaxies with stellar mass greater than $M_*$ at the same redshift, i.e., $n(>M_*, z)=\int_{M_*}^{\infty}\phi(M'_*, z)dM'_*$.

However, as can be seen in the upper panel of Figure \ref{fig:SMF}, the observed stellar mass function has mass-dependent completeness, especially at the low-mass end, arising from the complicated target selection and magnitude limit. We adopt a form of the complete SDSS stellar mass function
measured by \cite{2018ApJ...858...30G} (G18 for short) to make the galaxy-subhalo abundance matching. G18's mass function was reconstructed based on the galaxy clustering measurements. Then by comparing the observed stellar mass function with the complete function, we obtain the mass-selection function in each redshift bin and sample the mock galaxies in simulations according to this mass selection function.
By combining all redshift bins with $\Delta z=0.03$, we yield an all-sky mock LOWZ sample with the same angular selection and mass selection over the redshift range of (0.2, 0.43).  The radial selection function is found to be recovered accurately as shown in the bottom panel of Figure \ref{fig:SMF}. 

For the DR7 sample, we create volume-limited mocks in a similar way within the redshift range of $0.025<z<0.163$. We have created 108 mock DR7 survey catalogs in total extracted from the HR4 simulation. 
We select galaxies with $r$-band absolute magnitudes brighter than $-21.07$ after luminosity evolution correction is applied (see \cite{2007ApJ...658..884C,2010JKAS...43..191C} for details). 

\subsection{Multiverse Simulations}
\label{sec:multiverse}


\begin{table}
\centering
\caption{Matter density parameter at the present epoch and constant dark energy equation of state parameter of the Multiverse simulations. 
  \label{table: multiverse}}
\centering
\begin{tabular}{c|llll}
\hline
$w$\textbackslash $\Omega_m$ &0.21 &0.26 &0.31 &0.36\\
\hline
$-0.5$ &M1 &M3 \\
$-0.7$ &   &M4 \\
$-1$   &M2 &M5 &M7 &M9 \\
$-1.5$ &   &M6 &M8 &M10\\
\hline
\end{tabular}
\end{table}

The Multiverse simulations are a set of $N$-body simulations of the Cold Dark Matter universes with constant or dynamic dark energy.  All simulations share exactly the same random number sequence used to generate the initial conditions. Therefore, the amount of cosmic variance is exactly the same between simulations, and the difference is solely due to the cosmology adopted. The power spectrum of each model is normalized so that they produce the same rms of the matter fluctuation ($\sigma_8=0.794$) at $z=0$. These simulations have a periodic comoving box size of 1024 $h^{-1}{\rm Mpc}$ and $N_p=2048^3$ dark matter particles, which are designed for the study of the effects of cosmological parameters on matter/galaxy clustering, as well as its redshift evolution. These simulations allow us to make a proper comparison between models down to non-linear scales without the influence of cosmic variance. 

In Table \ref{table: multiverse}, we list ten simulations in the Multiverse simulation used in this paper. Only those with constant dark energy equation of state are used in this study. The simulation M5 adopts our fiducial model, the 5-year WMAP cosmology. For the other models the matter density parameter at $z=0$, $\Omega_m$, is varied from 0.21 to 0.36 with the step of 0.05  while the dark energy equation of state parameter $w$ ranges from $-0.5$ to $-1.5$. Multiverse simulations are geometrically flat universes and all the remaining cosmological parameters, $\Omega_b,h,\sigma_8$, and $n_s$, are the same as those of HR4.

Since these simulations have much smaller box size than HR4, we use a different approach to generate mocks.
We create the full-sky DR7 mocks with multiverse simulations by placing an observer at the center of the simulation box. To reduce the statistical uncertainty in the CF measurement, we do not adopt any angular mask. 
For higher redshift samples ($z_i, i\ge2$), we also choose to create mocks inside a cubic box instead of the way of spherical slice for the following reasons. 
First, the box size of multiverse simulation is too small to produce realistic mock catalogs. The mock created from the entire box allows us to fully utilize the large-scale structures in the simulation. Second, the Multiverse simulations are used only to estimate the differences of the systematics between different cosmological models and the absolute measurement is not needed. This is possible because all the Multiverse simulations are run with the same random seed. 

We also account for the effect from mass incompleteness  by using an average mass selection function for each redshift slice $z_i$.
To include the RSD effect, we treat the (comoving) $x_3$ axis of the simulation box as the LOS direction and use the following equation to shift the LOS coordinates of halos from real to redshift space:
\begin{equation}
\widetilde{x}_3 = x_3 + v_3\frac{1+z_i}{H(z_i)},
\end{equation}
where $\widetilde{x}_3$ denotes the shifted $x_3$-coordinate, $v_3$ is the proper peculiar velocity of galaxies in real space along $x_3$ direction, and $H(z)$ the Hubble parameter. We also generate two other catalogs by choosing $x_1$ and $x_2$ as the LOS direction respectively. As the last step, we average the CF measured from three mocks to reduce the statistical uncertainty. 

Although the mock construction for Multiverse simulations here is less realistic, it is used only to capture the CF shape difference between different cosmological models.
To estimate the full amount of the nonlinear systematics in the redshift evolution of CF, we will use the CFs of Multiverse simulations in combination with the CF measured from the HR4 mocks as described in \S\ref{sec:shape-evolution}.

\section{Method}
\label{sec:method}
\begin{figure*}[!htb]
    \centering
     \subfigure{
     \includegraphics[width=1\linewidth, clip]{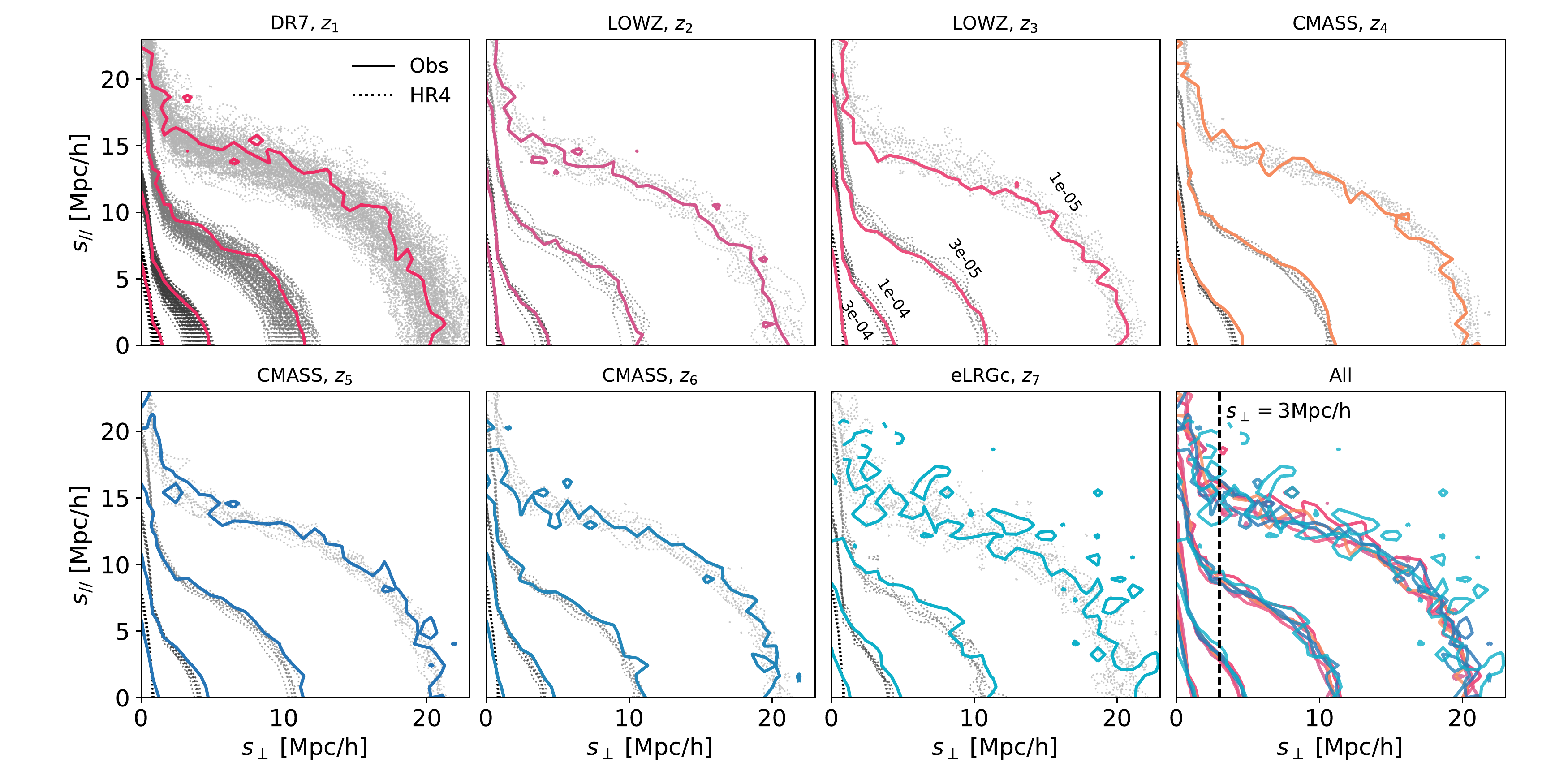}}
   \caption{Normalized two-point correlation functions of seven observational samples in increasing order of redshift from $z_1$ to $z_7$ (solid lines). 
   The dotted lines in grey are the measurements from HR4 mocks. All the observed correlation functions are superposed on one another in the last panel. Agreement between all the seven samples is excellent on the scales $s_\perp>3 h^{-1}$Mpc.
   A cosmology with $\Omega_m=0.26$ and $w=-1$ is used for the plots.}
    \label{fig:2D-7samples}
\end{figure*}

\begin{figure}[!htb]
    \centering
     \subfigure{
     \includegraphics[width=1\linewidth, clip]{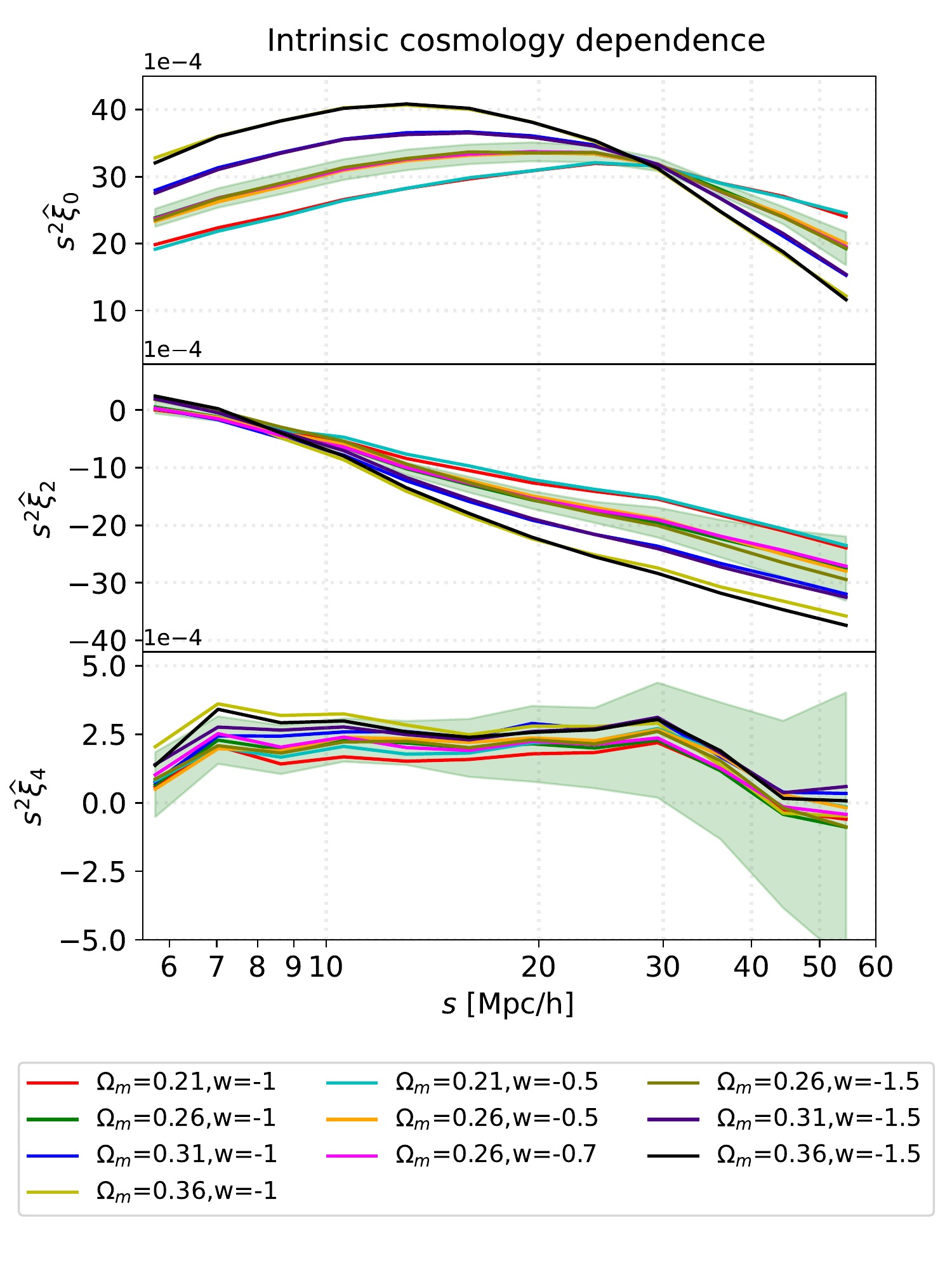}}
   \caption{Intrinsic shape moments of the normalized redshift-space CF obtained for 10 different  Multiverse simulation cosmologies in the second redshift slice $z_2$. Top: $\nxi_0(s)$, middle: $\nxi_2(s)$, bottom: $\nxi_4(s)$. Models with the same $\Omega_m$ nearly overlaps with one another. The shaded areas attached to the fiducial model show the $1\sigma$ uncertainties, which are estimated from the MultiDark mock simulations (see the text).}
    \label{fig:intrinsic}
\end{figure}

\subsection{Shape of the Two Point Correlation Function }
We use the two-point CF statistic to characterize the clustering of galaxies measured according to the Landy-Szalay estimator \citep{1993ApJ...412...64L}:
\begin{equation}
    \xi(r_\parallel,r_\perp)=\frac{{ DD-2DR+RR}}{{ RR}},
\end{equation}
where $DD, DR$, and $RR$ are the galaxy–galaxy, galaxy–random, and random–random pair counts, respectively. $r_\parallel$ is the pair separation parallel to the line of sight (LOS), and $r_\perp$ is the pair separation perpendicular to LOS in comoving coordinates. For a pair of galaxies at positions $\bf{r_1}$ and $\bf{r_2}$ with respect to the observer, we define:
\begin{equation}
    r_\parallel=\frac{\bf{r}\cdot \bf{\bar r}}{|\bf{\bar r}|}, \  r_\perp=\sqrt{\bf{r}\cdot \bf{r}-r_\parallel^2},
\end{equation}
where ${\bf{\bar r}}=({\bf{r_1}}+{\bf{r_2}})/2$, 
${\bf{r}}={\bf{r_2}}-{\bf{r_1}}$.
In polar coordinates, this relation can be described as $\mu=r_\parallel/|{\bf r}|=cos\ \vartheta$, where $\vartheta$ is the angle between the pair separation $\bf r$ and the line of sight. 

A pair of galaxies spanning $\Delta z$ in redshift and $\Delta\theta$ in angle have their comoving separations $r_\parallel=[c/H(z)]\Delta z$ and $r_\perp=(1+z)D_A(z)\Delta\theta$
along and across LOS, respectively. When a wrong cosmology is adopted in the conversion from redshift to comoving distance, they are incorrectly changed by the following factors, 

\begin{equation}
\frac{[r_\parallel]_{\rm wrong}}{[r_\parallel]_{\rm true}}=\frac{H_{\rm true}(z)}{H_{\rm wrong}(z)}, \ \frac{[r_\perp]_{\rm wrong}}{[r_\perp]_{\rm true}}=\frac{D_{A,{\rm wrong}}(z)}{D_{A,{\rm true}}(z)}.  
\end{equation}
The AP test exploits the fact that $r_\parallel$ and $r_\perp$ scale differently with the change of cosmology as in Equation (6). This results in apparent changes in shape, $r_\parallel / r_\perp$, or volume, $r_\parallel {r_\perp}^2$.

In our AP test we make use of the two-dimensional shape of CF \citep{2019ApJ...881..146P} in redshift space.
After we measure the CF, we normalize $\xi$ by the normalization factor which is a volume integral of $\xi$ up to radial separation of $s_{\rm max}$ in redshift space, 
\begin{equation}
    {\bar \xi}(z)=2\pi\int_0^1d\mu\int_0^{s_{\rm max}}s^2ds\xi(s,\mu,z).
\end{equation}
This is a form of the $J_3$ integral \citep{1980lssu.book.....P}.
$s_{\rm max}$ is chosen to be sufficiently large 
so that the normalization is insensitive to $s_{\rm max}$. Here we choose $s_{\rm max}=60\, h^{-1}{\rm Mpc}$.
An attendant benefit of this operation is that our results become independent of the overall evolution of clustering amplitude due to gravitational evolution and change of galaxy bias in observational samples.

Figure \ref{fig:2D-7samples} shows the normalized CF, $\nxi=\xi/{\bar \xi}$,
measured for seven redshift slices.
The fiducial cosmological model (M5 in Table \ref{table: multiverse}) is adopted in the analyses of observational samples.
The normalized CFs measured from the HR4 mocks (grey dotted contours) match observations (solid contours) quite well on most scales except at $s_\perp  \lesssim 2 h^{-1}$Mpc. At these very small separations we notice some differences between $\nxi^{\rm obs}$ and $\nxi^{\rm mock}$ that depend on scale and redshift. 

The last panel shows the CFs of the seven observational samples at different redshifts superposed on one another. It should be noted that, taking into account the statistical fluctuations, $\nxi^{\rm obs}(s_\parallel, s_\perp)$ have almost the same shape. 
Observational samples also show some differences of CF in different redshift slices on very small scales at $s_\perp  \lesssim 2 h^{-1}$Mpc. This can be related with the fiber collision and evolving Fingers of God.
We correct $\nxi^{\rm obs}$ for such systematic effects using mock galaxies. Readers are referred to Appendix \ref{sec:fiber} for details.  
Even after the correction, we exclude the region $s_\perp< 3 h^{-1}{\rm Mpc}$ of the 2d CF plane from our analysis to safely stay away from the scales that are affected by fiber collision and Fingers of God.

Following the procedure of \cite{2019ApJ...881..146P}, we compress the information in the shape of $\nxi$ into the Legendre multipole moments and study its angular and radial dependence:
\begin{equation}
\label{eq:legendre}
     \nxi(s,\mu,z)\approx\sum_{l=0,2,4}\nxi_l(s,z)P_l(\mu),
\end{equation}
where $P_0=1$, $P_2=(3\mu^2-1)/2$ and $P_4=(35\mu^4-30\mu^2+3)/8$. As we exclude the region with $s_\perp <3\, h^{-1}{\rm Mpc}$ of $\nxi$, we cannot use the decomposition formula $\nxi_l=\int\nxi P_l(\mu)d\mu$. Instead we use a $\chi^2$ fitting method \citep{2019ApJ...881..146P}. In the subsequent parts of the paper, we use $\nxi_{0,2,4} (s)$ to examine the shape evolution of the CF.

\subsection{Cosmology Dependence of the shape of CF}
\label{sec:nxi-sys}

\cite{2019ApJ...881..146P} has found that the intrinsic gravitational evolution of $\nxi_{0,2,4}(s)$ is cosmology-dependent. Although the effect of cosmology on the shape of CF is tiny compared to the full RSD effect dominated by peculiar velocities, it can be comparable to the geometrical distortion when a sub-percent level uncertainty is required for cosmological parameters and therefore needs to be estimated as accurately as possible.
Figure \ref{fig:intrinsic} shows $\nxi_{0,2,4}(s)$ measured for 10 multiverse simulations listed in Table 2. 
$\nxi_0$ and $\nxi_2$ are of similar order, and their cosmology dependence is accurately measured from our Multiverse simulations. In comparison, $\nxi_4$ is an order-of-magnitude smaller, and its cosmology dependence is weaker.

The uncertainty ranges of those moments are calculated from MultiDark mocks \citep{2016MNRAS.456.4156K}. The MultiDark mocks are produced with the PATCHY code, by generating a dark matter field and using biasing prescriptions to populate it with mock galaxies. The mocks have been calibrated using a reference galaxy catalogue based on the halo abundance matching modelling of the BOSS DR11\&DR12 galaxy clustering data and on the data themselves.

\subsection{Redshift Evolution of the shape of CF}
\label{sec:shape-evolution}
Let us denote the difference of the normalized CFs between two redshifts as 
\begin{equation}
    \Delta\nxi(z_i,z_j)=\nxi(z_i)-\nxi(z_j).
\end{equation}
In the likelihood estimation below, we use the data points of $\Delta\nxi$ within the radius range of $s =6.3 - 15 h^{-1}{\rm Mpc}$. Compared to the effective radius range of $5-15 h^{-1}$ Mpc for cosmological constraints suggested by \cite{2019ApJ...881..146P}, we use a slightly larger lower bound on the radius given that our galaxy samples are more sparse.

The evolution in the intrinsic shape of redshift-space CF between redshift slices in a particular cosmology is estimated from 
\begin{equation}
\begin{split}
\Delta\nxi^{\rm{sys}}(\Omega_m^M,w^M)=&\Delta\nxi^{M}(\Omega_m^M,w^M)-\Delta\nxi^{M}(\Omega^{M5}_{m},w^{M5}) +\\
&\Delta\nxi^{H}(\Omega^H_{m},w^H),
\end{split}
\end{equation}
where $M$ refers to a multiverse simulation cosmology, and $\Omega^{M5}_{m}=\Omega^{H}_m =0.26, w^{M5}=w^H=-1$, which are the parameters of the fiducial model or HR4 cosmology (denoted by `H'). This means that accurate measurement of the systematic shape evolution is made in HR4, and the Multiverse simulations are used to extrapolate the estimation to a particular cosmology using only the difference relative to the fiducial model. Thanks to its larger box size, more mock surveys can be made within HR4 and the cosmic variance can be reduced in the estimation of the systematic evolution.

We then use a second-order polynomial 
\begin{equation}
    s^2\Delta\nxi_k^{\rm sys}\approx\alpha_{k1}^{\rm sys}+\alpha_{k2}^{\rm sys}{\rm log}(s)+\alpha_{k3}^{\rm sys}{\rm log}(s)^2,
\end{equation}
to fit the $s$-dependence of $k$-th moment in the Legendre polynomial expansion of CF (see Eq. \ref{eq:legendre}) shown in Figure \ref{fig:intrinsic}. Once all the coefficients $\alpha^{{\rm sys}}_k(\Omega_m^M,w^M)$ are calculated for ten cosmological models, the coefficients for any neighboring cosmologies are interpolated or extrapolated using a third-order polynomial. 
Here we choose to use a third-order polynomial as it is found to work much better than the lower order polynomials, and the results no longer changes when further using a fourth-order polynomial.

\subsection{Likelihood Analysis with Mock Data}
\label{sec:likelihood}
\begin{figure}[!htb]
    \centering
     \subfigure{
     \includegraphics[width=1\linewidth, clip]{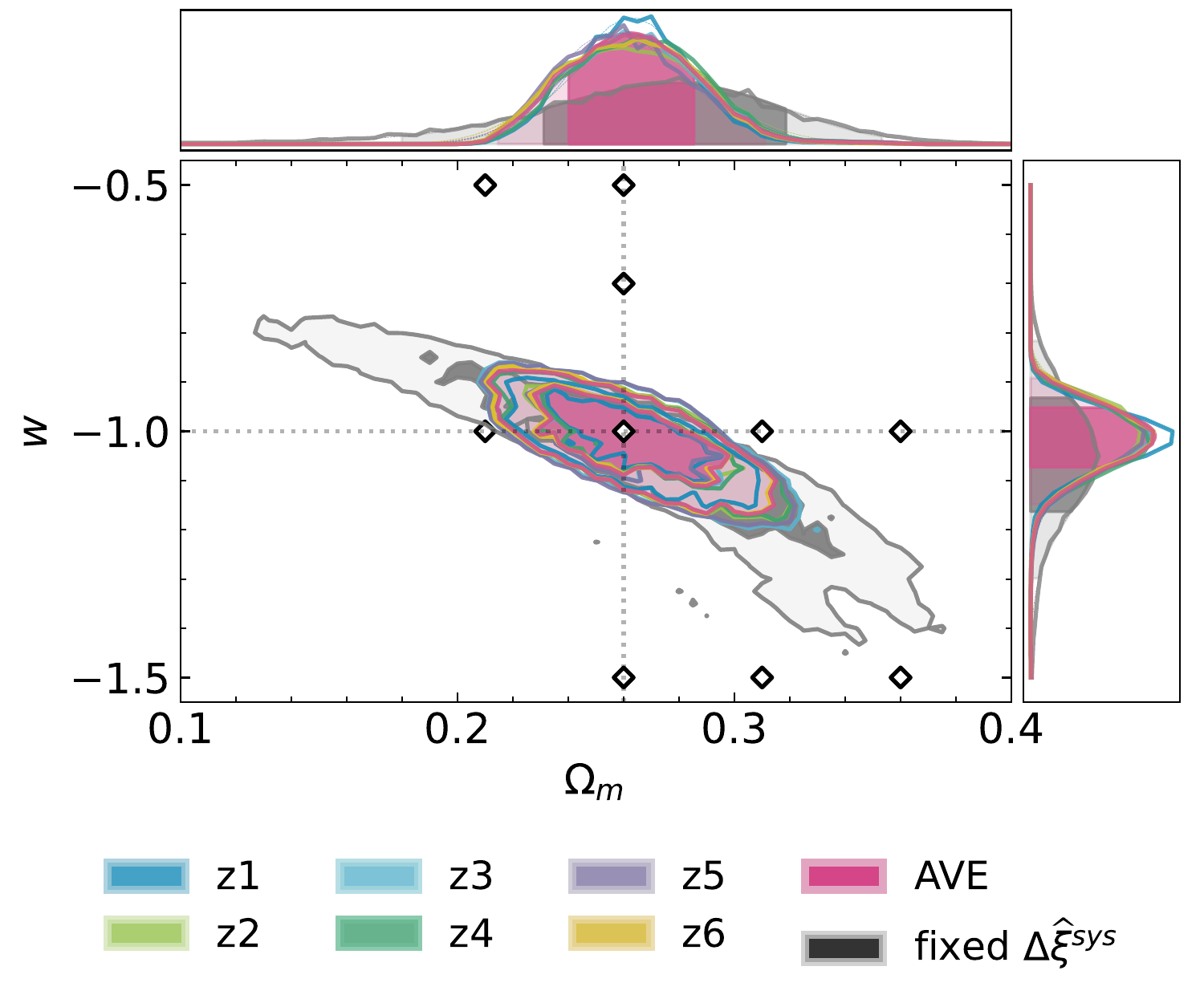}}
   \caption{Likelihood function maps $\mathcal{L}(\Omega_m,w)$ from our extended AP test analysis using the baseline mock samples.
   The contours in different colors correspond to the cases when the slice at $z_i$ is chosen as  
   the reference for measuring the relative CF shape evolution across redshift slices.  Cosmology-dependent systematic corrections to the shape evolution are made ($\Delta\nxi^{\rm sys}(\Omega_m,w)$).  We average over all choices of the reference slice for the final constraint (pink color). For comparison, the constraint assuming 
   fixed amount of systematics obtained for the fiducial cosmology ($\Delta\nxi^{\rm sys}=\Delta\nxi_{\rm fid}(\Omega_{m},w)$) is shown with grey color contours, which is more stretched in the plane.  The diamond symbols denote the ten cosmologies with varying $\Omega_m$ and $w$ covered by the Multiverse simulations given in Table 2. The sub-panels present marginalized constraints for each parameter with curves enclosing the 1$\sigma$ and 2$\sigma$ regions. 
   }
    \label{fig:mock_cc}
\end{figure}

In this work we use realistic mock survey samples that incorporate all the important aspects of observation such as survey angular and radial selection functions and galaxy mass sampling rate variation in order to test our method.

We will calculate the likelihood of various cosmological models when the true cosmology is that of HR4. The mock survey samples drawn from HR4 are used in this calculation.  
For a particular cosmology adopted we first transform the redshifts of galaxies in the mock samples to comoving distances, and measure the normalized CF of galaxies, 
$\nxi(\Omega_m,w,z)$, 
at seven redshift bins. Then we choose a reference redshift, $z_{\rm ref}$, and measure the CF shape difference between $z_i$ and $z_{\rm ref}$
\begin{equation}
\Delta\nxi(z_i,z_{\rm ref})=\nxi(z_i)-\nxi(z_{\rm ref})-\Delta\nxi^{\rm{sys}}(z_i,z_{\rm ref}), 
\end{equation}
where $z_i\neq z_{\rm ref}$. We also make a correction of the difference for the intrinsic shape evolution due to gravitational evolution, survey sample variation, and various observational effects.
$\Delta\nxi^{\rm{sys}}$ is obtained as described in the previous section.

We perform a joint analysis taking into account the correlations between different $z_i$ bins for a fixed $z_{\rm ref}$. The likelihood for any adopted cosmology and the reference redshift $z_{\rm ref}$ chosen is given by $\mathcal{L}_{z_{\rm ref}}={\rm exp}(-\chi^2_{z_{\rm ref}}/2)$, where
\begin{equation}
\label{eq:chi2}
    \chi^2_{z_{\rm ref}}(\Omega_m,w)=\sum_{k=0,2,4}\sum_{\alpha,\beta} P_k^{z_{\rm ref}}(s_\alpha)\cdot (C^k_{\alpha\beta})^{-1} \cdot P_k^{z_{\rm ref}}(s_\beta).
\end{equation}
Here $P_k^{z_{\rm ref}}=(...,\Delta\nxi_k(z_i,z_{\rm ref}),...,\Delta\nxi_k(z_j,z_{\rm ref}),...)$ ($z_i \neq z_{\rm ref}\neq z_j$, $i<j$), $s_\alpha$ is the $\alpha$-th pair separation bin radius, and $C^k_{\alpha\beta}$ the covariance matrix of the $k$-th moment in the Legendre polynomial expansion.

To assess the CF covariance we need a large number of mock samples. We use the MultiDark PATCHY mock catalogs \citep{2016MNRAS.456.4156K} to make mocks for the BOSS DR12 data, and EZmock catalogs \citep{2021MNRAS.503.1149Z} for the eBOSS LRG data. Both PATCHY and EZmock generate the dark matter density field using the Zel'dovich approximation \citep{1970A&A.....5...84Z}. The PATCHY mocks have been calibrated using a reference simulation to obtain a detailed galaxy bias evolution spanning the redshift range from 0.15 to 0.75. On the other hand, the EZmock catalogs have been calibrated directly with the auto correlations of data in different redshift bins. For the DR7 sample, we use 108 mocks from the HR4 $N$-body simulation to assess the covariance.

We examine the 2D parameter space of the matter density parameter $0.1<\Omega_m<0.5$ and dark energy equation of state $-1.5<w<-0.5$ with intervals of $\Delta \Omega_m=0.005$ and $\Delta w=0.025$. We measure the CF of the mock galaxies for the cosmology corresponding to each grid point in the ($\Omega_m,w$) space. The likelihood maps in the $\Omega_m-w$ plane derived from the above analysis are shown in Figure \ref{fig:mock_cc}. Note that the true cosmology has ($\Omega_m,w)=(0.26,-1)$. The 68\% and 95\% probability contours are shown. 

When the likelihood is calculated, one can choose any of the survey sample redshifts as the reference $z_{\rm ref}$. In Figure \ref{fig:mock_cc} contours in different color show how much the likelihood function can vary as the reference epoch is varied.
Overall, these contours are quite consistent with each other in both size and shape. This demonstrates that our results are insensitive to the choice of $z_{\rm ref}$.
We take the reference redshift-averaged likelihood as the final result (pink contours), $\mathcal{L}=\sum_{i=1}^n\mathcal{L}_{z_i}/n$, where $n=6$ and $\mathcal{L}_{z_i}$ is the likelihood when $z_i$ is the reference redshift.

To make evident the benefit of using cosmology-dependent correction for systematic shape evolution, we also present the result derived using a fixed correction $\Delta\nxi^{\rm sys}_{\rm fid}(\Omega_m^{M5},w^{M5})$ (grey color contours). The likelihood contours with fixed systematics correction are peaked basically at the same place but are much broader in the parameter space compared to the model-dependent correction case. Model-dependent systematics correction reduces 
the statistical error in $\Omega_m$ by a factor of 1.8 and 
the error in $w$ by a factor of 2, thus contributing to break the parameter degeneracy.  
The benefit comes from the fact that
the shape of CF changes as $\Omega_m$ changes as can be seen in Figure 4, which is taken into account in this work. This makes the likelihood contours much shorter along the $\Omega_m$ direction.

\section{Result}
\label{sec:result}
\begin{figure*}
    \centering
     \subfigure{
     \includegraphics[width=0.8\linewidth, clip]{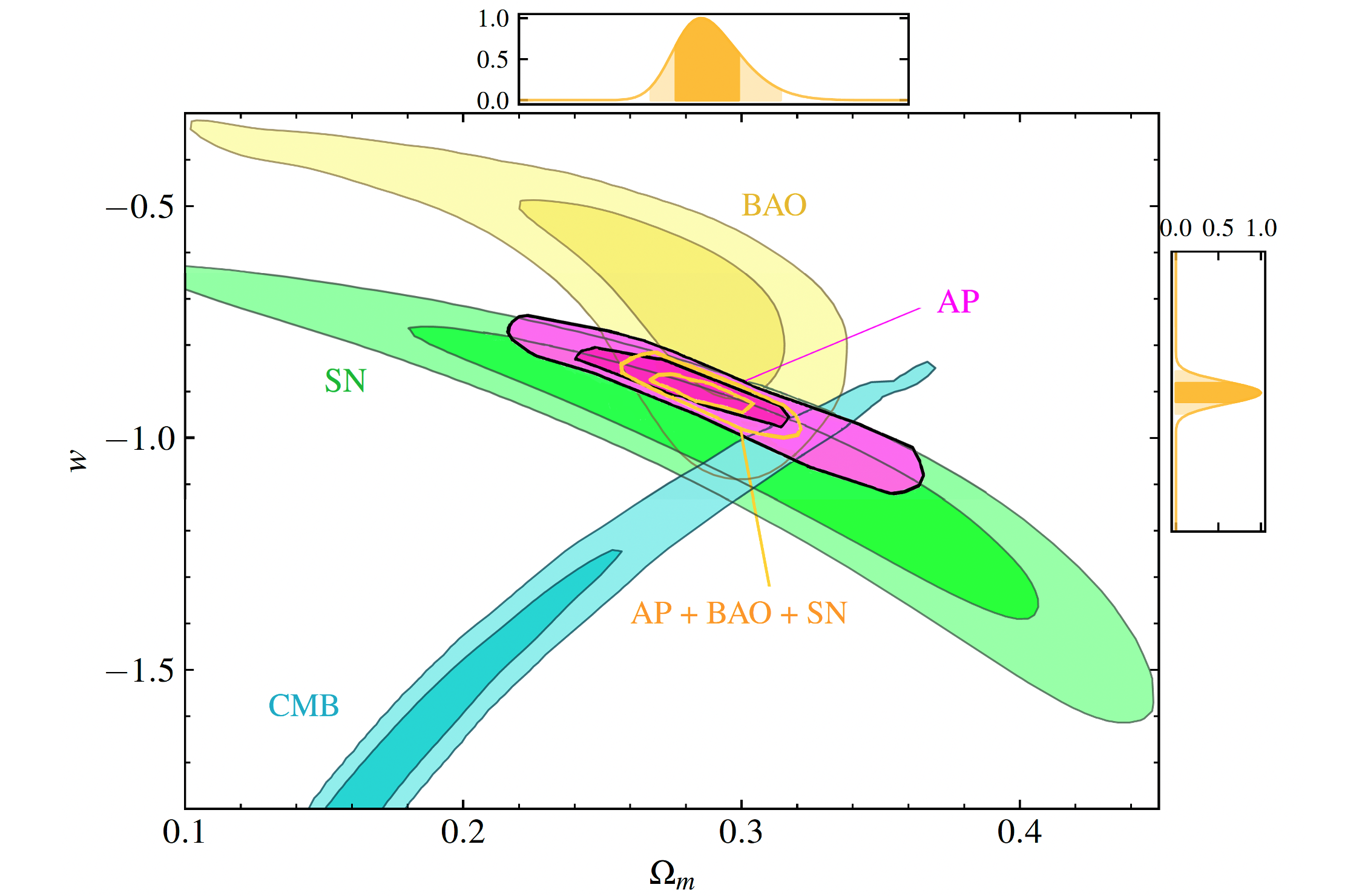}}
   \caption{Likelihood function maps in the $w-\Omega_m$ plane from our extended AP test using the baseline SDSS data. We also show constraints from BAO/SN/CMB probes for comparison. 
   The contours in gold shows the likelihood function of our AP test combined with two other low-redshift probes. The sub-panels show the marginalized probability distribution function of each cosmological parameter.
   The likelihood functions of other cosmological probes are borrowed from literature \cite{2021PhRvD.103h3533A}; the SDSS BAO-only measurements \citep{2015MNRAS.449..848H,2017MNRAS.470.2617A,2021MNRAS.500.3254R,2021MNRAS.501.5616D}, Pantheon SNe I$a$-only measurements \citep{2018ApJ...859..101S} and Planck CMB-only measurements \citep{2020A&A...641A...1P}.
   }  
    \label{fig:result_all}
\end{figure*}

\begin{figure}[!htb]
    \centering
     \subfigure{
     \includegraphics[width=1\linewidth, clip]{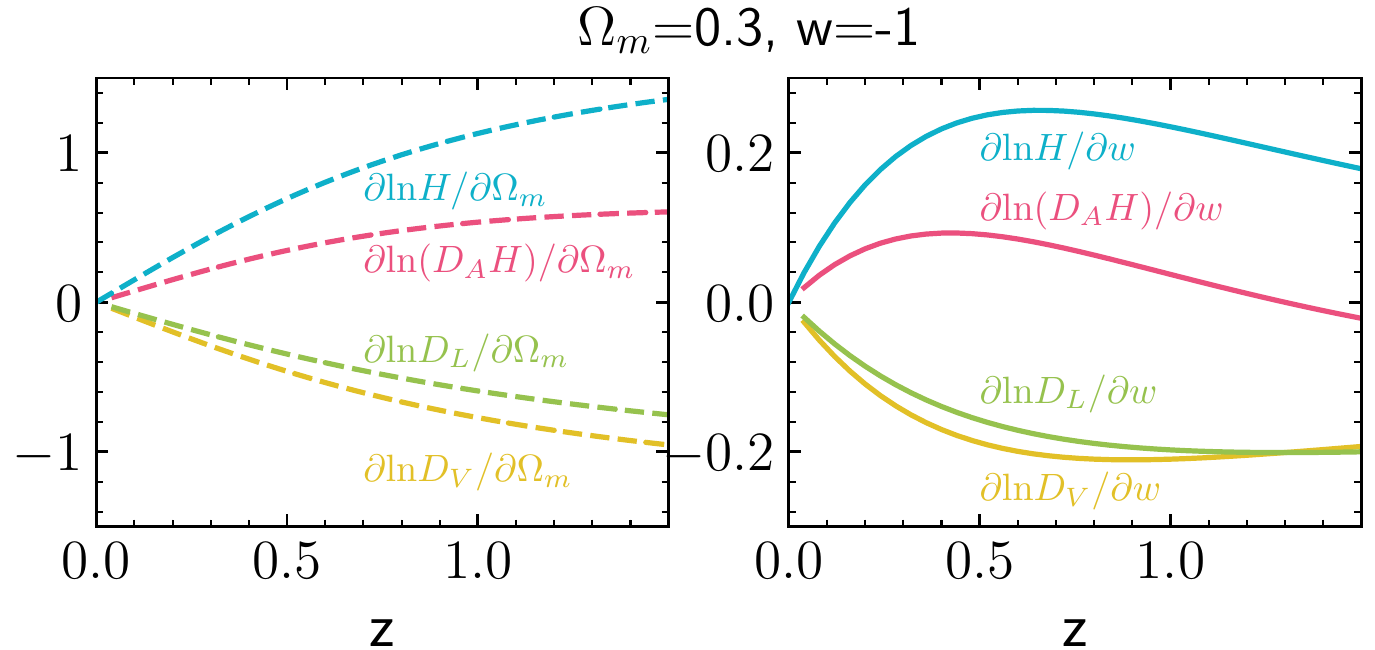}}
   \caption{The sensitivity of $D_A H$ to $\Omega_m$ and $w$, quantified by $\partial{\rm ln}(D_A H)/\partial\Omega_m$ (left panel) and $\partial{\rm ln}(D_A H)/\partial w$ (right panel). Sensitivity to $w$ decreases at high  redshifts and crosses zero at $z\sim1.3$. The results for $H(z)$, $D_L$ and $D_V$ are plotted for comparison. A cosmology of $\Omega_m=0.3$ and $w=-1$ is used.}
    \label{fig:senst}
\end{figure}

\begin{figure}[!htb]
    \centering
     \subfigure{
     \includegraphics[width=1\linewidth, clip]{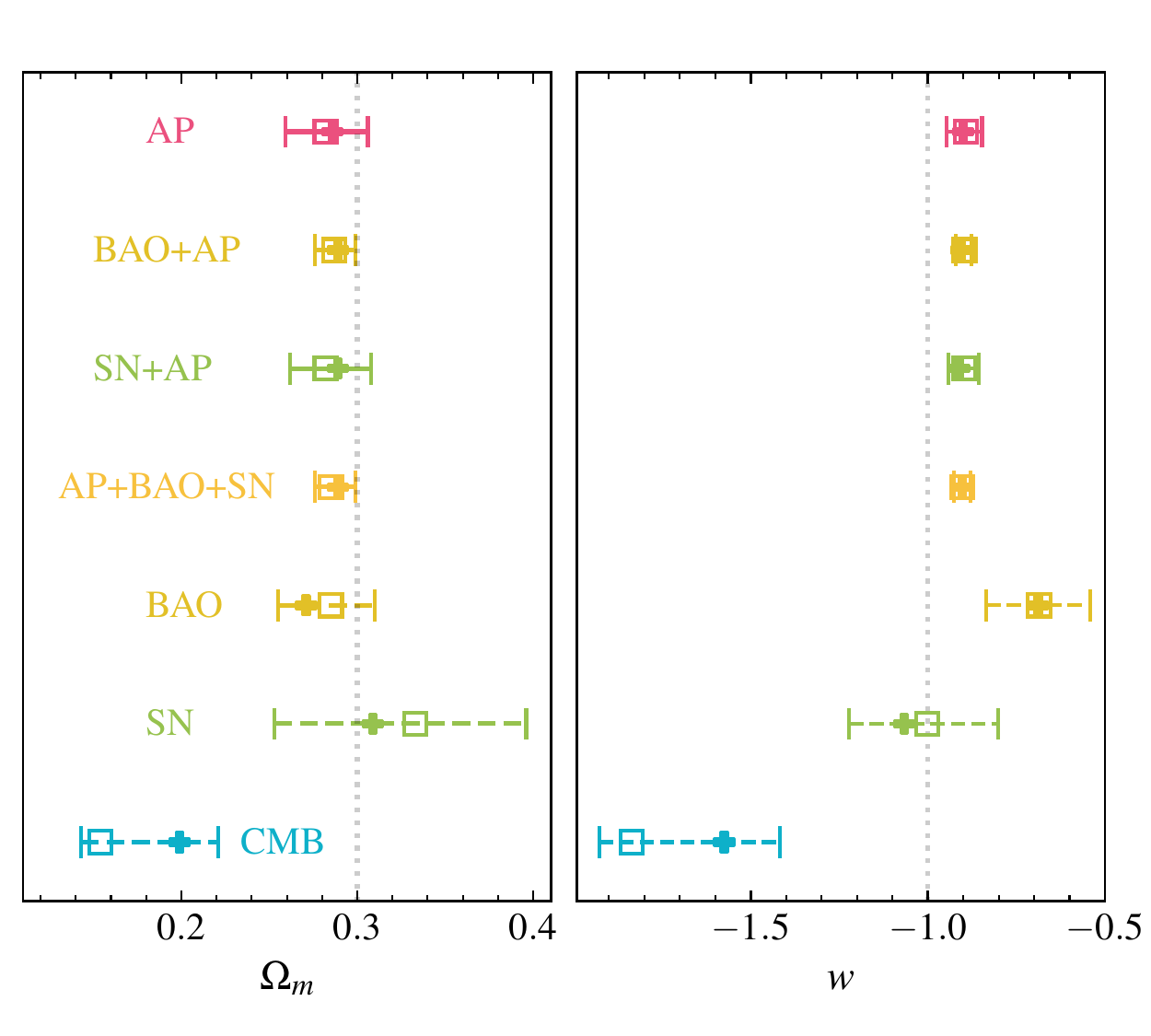}}
   \caption{Improvement in $w$CDM constraints by including the AP measurement. Results are shown for each data set combination presented in the text and Table 3. Squares are the ``peak" values, which are the parameter values at the highest probability in maximum likelihood estimate, and crosses are the``mean" values referring to the average value of the parameter. The improvement factor in parameter uncertainty is 3-7.}
    \label{fig:woLRG}
\end{figure}

\begin{table*}
\centering
\caption{
Marginalized best-fit value, $\Omega_m$ and $w$, average value, $\langle\Omega_m\rangle$ and $\langle w\rangle$, and 68$\%$ confidence limits, $\sigma(\Omega_m)$ and $\sigma(w)$, estimated from various cosmological probes and their combined likelihood analyses.
The baseline samples include the six low redshift SDSS slices while all samples include the eBOSS LRG sample as well. The ‘joint’ constraint considers the cross-correlations between different radial modes (see the text).
Our main result is from combination of baseline AP$+$BAO$+$SN.
  \label{Table-result}}
\begin{minipage}{180mm}
\centering
\begin{tabular}{cc|ccc|ccc}
\hline
\hline
galaxies&probes&$\Omega_m$  & $\langle\Omega_m\rangle$ & $\sigma(\Omega_m)$ &$w$ & $\langle w \rangle$ & $\sigma(w)$\\
\hline
                 &BAO           & $0.285_{-0.030}^{+0.025}$
                 &0.271 &0.036 &$-0.686_{-0.149}^{+0.144}$&$-0.689$ &0.147\\
                 &SN            & $0.333_{-0.080}^{+0.063}$&0.309 &0.076 &$-1.0024_{-0.22}^{+0.2}$&$-1.066$ &0.216\\
                 &CMB           & $0.154_{-0.011}^{+0.067}$&0.199 &0.049 &$-1.836_{-0.092}^{+0.419}$&$-1.575$ &0.269 \\
baseline         &AP            & $0.282_{-0.023}^{+0.024}$&0.286 &0.025 &$-0.892_{-0.050}^{+0.045}$&$-0.900$ &0.050\\
baseline         &AP(joint)     & $0.276_{-0.021}^{+0.024}$&0.280 &0.024 &$-0.892_{-0.045}^{+0.04}$&$-0.899$ &0.047\\
baseline         &AP+BAO        & $0.287_{-0.011}^{+0.012}$&0.289 &0.013 &$-0.897_{-0.025}^{+0.020}$&$-0.905$ &0.025\\
baseline         &AP+SN         & $0.282_{-0.020}^{+0.026}$&0.289 &0.024 &$-0.897_{-0.045}^{+0.040}$&$-0.911$ &0.049\\
{\bf baseline}         &{\rm\bf AP}+{\rm\bf BAO}+{\rm\bf SN}         & $0.285_{-0.009}^{+0.014}$&0.289 &0.012 &$-0.903_{-0.023}^{+0.023}$&$-0.903$ &0.023\\
baseline         &AP+CMB        & $0.317_{-0.005}^{+0.008}$&0.320 &0.007 &$-0.982_{-0.025}^{+0.020}$&$-0.989$ &0.026\\
all              &AP            & $0.255_{-0.023}^{+0.023}$&0.255 &0.023 &$-0.842_{-0.05}^{+0.05}$&$-0.844$ &0.054\\
\hline
\end{tabular}
\end{minipage}
\end{table*}

We apply our method to the seven redshift samples of the SDSS survey shown in Figure 1, following the likelihood analysis described in \S\ref{sec:likelihood}.  
We adopt the geometrically-flat $w$CDM cosmology with two free parameters, the constant dark energy equation of state $w$ and the matter density parameter at the present epoch $\Omega_m$ as our main goal is test if the flat $\Lambda$CDM model is consistent with observation or not.
We first present the result for the baseline galaxy samples of six lower redshift slices. The constraints are shown by the magenta contours in Figure \ref{fig:result_all}.

In the figure 
we compare the constraints on $w$ and $\Omega_m$ from our AP test with those from three other probes\footnote{We use the MCMC chains and likelihoods from the
https://www.sdss.org/science/cosmology-results-from-eboss\citep{2021PhRvD.103h3533A}.}; the SDSS BAO-only measurements (\cite{2015MNRAS.449..848H,2017MNRAS.470.2617A,2021MNRAS.500.3254R,2021MNRAS.501.5616D}; yellow contours), Pantheon SNe I$a$-only (SN for short) measurements (\cite{2018ApJ...859..101S}; green contours) and Planck CMB-only measurements (\cite{2020A&A...641A...1P}; blue contours).

We summarize all the single and combined constraints in Figure \ref{fig:senst} 
and Table \ref{Table-result}. The inclusion of AP alleviates the tension in $w$ between BAO and SN. The value of $w$ estimated from AP+BAO ($-0.897_{-0.025}^{+0.02}$) is quite consistent with that of AP+SN ($-0.897_{-0.045}^{+0.04}$) while the BAO-only result ($-0.686_{-0.149}^{+0.144}$) somewhat differs from the SN-only result ($-1.002_{-0.22}^{+0.2}$). The same is true for $\Omega_m$. Compared to BAO alone, AP+BAO improves the precision of $\Omega_m$ by a factor of 2.8 and $w$ by a factor of 5.9.  For AP+SN, the improvement factors are 3.2 and 4.4 respectively. For AP+CMB, the factors are $\sim7$ for both parameters.  Nonetheless, obvious discrepancies are found between CMB and the other three low redshift probes. 
Therefore, regardless of the benefit of AP+CMB in reducing the uncertainties $\sigma_{\Omega_m}$ and $\sigma_w$, such combinations are expected to be statistically meaningless, as they barely overlap in the parameter space.

Figure \ref{fig:result_all} tells that there will be little improvement in parameter constraining when SNe I$a$ result is combined to our AP test because its constraint is relatively much weaker and degeneracy direction is nearly parallel to that of the AP test.
On the other hand, the constraint from BAO, also being relatively much weaker, is found to be useful as its likelihood contours have different slope. When the BAO result is combined with our AP test result, both $\sigma_{\Omega_m}$ 
and $\sigma_w$ 
are found to be reduced by factors of 2, even though both methods used 
almost the same survey catalogs \footnote{The SDSS BAO analysis used galaxy and quasar samples from all the SDSS surveys at $z<2.2$ and Ly$\alpha$ forest observations over $2<z<3.5$.}. 
It should be noted that our AP test places constraints on cosmological parameters governing the expansion history of the universe much tighter than those from the BAO method. 
The reason is that the BAO method uses the BAO feature in CF on scales greater than $100 h^{-1}{\rm Mpc}$ while our AP method uses the shape information of CF around $10 h^{-1}{\rm Mpc}$. Our AP test uses much stronger clustering signal and the statistical sample size is effectively about 1000 times larger \citep{2019ApJ...881..146P}.

As in Figure \ref{fig:mock_cc}, the AP test results in a degeneracy between $\Omega_m$ and $w$ with a negative slope.
This can be qualitatively understood as follows. The AP effect depends on the ratio of the observed size across the line of sight to radial size,  $\Delta r_\perp/\Delta r_\parallel\propto D_A H$.
The product $D_A H$ is an increasing function of both $\Omega_m$ and $w$. The pink curves of Figure 7 
demonstrate that the slope of the function with respect to $\Omega_m$ or $w$ is positive well beyond $z=1$ in the case of a cosmology with $\Omega_m= 0.3$ and $w=-1$. Therefore, if $\Omega_m$ is increased and $w$ is decreased or vice versa, $D_A H$ can remain the same and the AP test cannot distinguish between the two cosmologies. Figure 8
also shows that the dependence of $D_A H$ on $w$ weakens at $z\gtrsim 1$ while the $\Omega_m$ dependence monotonically increases. This makes the direction of parameter degeneracy change as redshift increases. Therefore, to break the degeneracy one needs to either use additional probes that depends on $D_A$ and $H$ differently or employ a set of observational samples reaching redshifts greater than 1.

The degeneracy direction of the constraint from a probe in the $w$-$\Omega_m$ plane can be denoted as $\delta w\approx c\delta\Omega_m$. In the case of the AP test, the slope is given by $c_{\rm AP}=[\partial ({D_A H})/\partial\Omega_m]/[\partial ({D_A H})/\partial w$]. 
We find $c_{\rm AP}<0$ at $z< 1.3$ and $c_{\rm AP}>0$ at higher redshifts. Hence our AP analysis here yields a negative degeneracy direction as the SDSS samples used are at $z<0.8$. 

The degeneracy direction for Type I$a$ SNe is also negative in this parameter space. This is because the luminosity distance is a decreasing function of both $\Omega_m$ and $w$, as can be seen in Figure \ref{fig:senst}. Therefore, similarly to the AP test, when $\Omega_m$ and $w$ vary in opposite directions, $D_L$ can remain the same and the degeneracy direction becomes almost parallel to that of the AP test when low redshift samples ($z \lesssim 1$) are used. 
Even though they have almost the same degeneracy direction at low redshifts, in comparison with the SN probe, the AP test breaks its own degeneracy as the sample depth increases beyond $z=1$.

On the other hand, the geometrical degeneracy of the angular size of CMB anisotropies yields a degeneracy along the positive direction in the $w$-$\Omega_m$ plane \citep{1997MNRAS.291L..33B,1997ApJ...488....1Z}.
The angular size
depends 
on the angular diameter distance to the decoupling surface, which is a decreasing function of both $w$ and $\Omega_m$ in flat $\Lambda$CDM universes.
But the acoustic scale in the angular power spectrum depends mainly on the combination $\Omega_m^{1/2} D_A$, which increases with $\Omega_m$ but is a decreasing function of $w$. This is why the degeneracy direction for CMB has a positive slope in Figure 6.

The LRG sample of the eBOSS survey has much lower galaxy number density relative to those of other samples, and is not included in our analysis so far. 
However, it is still worthwhile to see if the LRG sample ($0.6<z<0.8$) helps to improve the AP constraint. We find that the inclusion of LRG sample slightly shifts the contour along the degeneracy direction towards lower $\Omega_m$ and less negative $w$ values ($\Omega_m=0.282_{-0.023}^{+0.024}\rightarrow0.255_{-0.023}^{+0.023}$, $w=-0.892_{-0.055}^{+0.045}\rightarrow-0.842_{-0.05}^{+0.05}$).  
Overall, the LRG sample does not contribute to the AP parameter constraint, most likely due to the large statistical uncertainties in the measurement of $\nxi$ introduced by the  low galaxy number density of the sample.

One may wonder whether or not the joint analysis of three radial modes expanded in Legendre polynomials affect the results.  To examine this possibility, we employ the full covariance matrix by including auto-correlations, ($P_0-P_0$, $P_2-P_2$, $P_4-P4$), and three cross-correlations, ($P_0-P_2$, $P_2-P_4$, $P_4-P_0$).
Then the constraints become $\Omega_m=0.276_{-0.021}^{+0.024}$ and $w=-0.892_{-0.045}^{+0.04}$, which shows that there is little change in the precision of parameter estimation in practice.

\section{Summary and Discussion}
\label{sec:conclusion}
We apply an extended version of the AP test using redshift evolution of the redshift-space CF to the current SDSS spectroscopic survey data. To this end, we measure the shape of the redshift-space two-point CF of galaxies within the redshift range of $0.025<z<0.8$. The galaxy number density of the SDSS DR7 and BOSS samples and their large survey areas meet our needs for accurately measuring the shape evolution of CF. We also employ a set of cosmological simulations with different cosmological parameters to estimate various systematics arising from intrinsic shape evolution. 
When the AP test is applied to constrain the $w$CDM models,
we obtain constraints on 
$w=-0.892_{-0.050}^{+0.045}$ and $\Omega_m=0.282_{-0.023}^{+0.024}$
from our AP test alone, which 
are much tighter than those from BAO or SN methods. While the SN result does not much improve the constraints when combined with our AP results, BAO is found to reduce the uncertainties by a factor of about 2. 

We find that our extended AP test combined with BAO and SN probes constrains the dark energy equation of state parameter to be 
$w_{\rm AP+BAO+SN}=-0.903_{-0.023}^{+0.023}$. This result is $4.2\sigma$ away from the standard concordance cosmological model with the cosmological constant $\Lambda$ or $w=-1$.
When our AP test is combined with Planck CMB result, we obtain
$w_{\rm AP+CMB}=-0.982_{-0.025}^{+0.020}$, which is consistent with $\Lambda$CDM.
This is because the low redshift cosmological probes such as standard shape (AP test), SN, and BAO jointly suggest the $w>-1$ universe while the Planck CMB prefers $w<-1$ universe.
Only when these two are combined, we find consistency with the `concordance' value $w=-1$.

The low redshift cosmological probes such as AP test, SN, and BAO have not been powerful enough to reveal the discrepancy with the CMB observation, which has put unprecedented tight constraints on various cosmological parameters. However, the late time data is now in a position to challenge it as galaxy redshift survey data increases and the low-redshift cosmological probes are elaborated to be more powerful, and our result gives evidence for $w>-1$ solely from low redshift probes .


\cite{2016ApJ...832..103L} has obtained the constraints $w = -1.07 \pm 0.15$ and $\Omega_m = 0.290 \pm 0.053$ from their AP test alone using the observational samples corresponding to the baseline samples in this work. It should be noted that there have been many important improvements in the data analysis here. Among them three are important. First, in this work we measure the mass selection function of galaxies at each narrow redshift bin, and use it when the abundance matching is made to generate mock samples from simulations while \cite{2016ApJ...832..103L} simply applied a sharp mass cut to simulated galaxies. As a result the mock galaxies in \cite{2016ApJ...832..103L}  had bias factors significantly higher than those of observed galaxies. Even though our AP method is not affected by the amplitude difference of CF, a small change in the shape of CF can be accompanied by the large difference in bias factor. Second, \cite{2016ApJ...832..103L}  applied the same correction for the systematics measured for the fiducial model (i.e. the flat $\Lambda$CDM model with $w=-1$ and $\Omega_m=0.26$ ) to all cosmological models in the likelihood analysis. As demonstrated in section \ref{sec:likelihood} and Figure \ref{fig:mock_cc}, this degrades the constraining power of the extended AP test. In this work we employed 10 Multiverse simulations combined with the HR4 simulation to accurately estimate the dependence of systematics on cosmology and nonlinear gravitational evolution. Third, we use the upgraded AP method developed by \cite{2019ApJ...881..146P}, where the full shape of CF is used in the AP test while Li et al.'s method uses only the angular shape of CF. Park et al. reported that there is about 40\% reduction in the uncertainties in parameter estimation due to the improvement.

Although our AP test method demonstrates its ability to accurately probe the expansion history of the universe, there are still possibilities for improvement. Here we list a couple of them.

(\romannumeral1) Deeper observational samples. The role of dark energy in driving cosmic expansion becomes less important at redshifts higher than about 1. Therefore, deep observational samples reaching beyond $z=1$ will allow the AP test to give tighter cosmological constraints with much less degeneracy.  
%

(\romannumeral2) Accurate estimation of systematic effects. Even though our AP test is relatively insensitive to nonlinear systematics effects, it is still very important to estimate them accurately in order to constrain cosmological parameters near or below a percent level and to extend the test to wider parameter space. The most interesting direction for accurate understanding the systematics is to use large cosmological gravitational and hydrodynamic simulations with realistic astrophysical processes included. 
This will allow us to understand how the CF on large scales is affected by various small-scale nonlinear effects such as galaxy feedback and mergers, and galaxy density and velocity biases among others.

The ongoing Dark Energy Spectroscopic Instrument (DESI) survey will provide us with data sets that are deeper, wider, and denser than the eBOSS LRG, ELG, and quasar samples. It will measure spectra of more than 30 million galaxies and quasars over 14,000 square degrees of the sky during its 5-year observing session. It is expected that the AP test will put tighter constraints on the expansion history of the universe with DESI.

The $\Lambda$CDM model with $\Omega_m\sim 0.3$ is known to have two critical problems. First, the value of vacuum energy density is much smaller than the theoretically motivated one by a factor of $\sim10^{54}$ as the electroweak or super-symmetry breaking at Tev gives $\rho_{\rm vac} > (1\ {\rm Tev})^4$ while the model has $\rho_{\rm vac}\sim (10^{-2} {\rm eV})^4$  \citep{2017AdSpR..60..166A}. Second, there is no natural explanation on the coincidence that $\Omega_m$ and $\Omega_{\Lambda}$ are of the same order at the present epoch.  The quintessence model \citep{1988ApJ...325L..17P}  (Smer-Barreto, V., \& Liddle, A. R. 2017, JCAP, 1, 023). was proposed to overcome these problems.  However, until now the statistical constraints on cosmological parameters have not been tight enough to judge whether or not the flat $\Lambda$CDM is inconsistent with observations while other models with mild dark energy dynamics are still consistent \citep{2022MNRAS.513.5686C}. Now observations seem to indicate that the dark energy may not be the cosmological constant but a dynamical quantity with $w>-1$ at $z<0.8$. This finding inevitably leads us to pay attention to the quintessence model. It remains to confirm our finding with upcoming larger observational data.

\section*{Acknowledgments}
The authors are grateful to the useful discussions with Ashley J. Ross and Anand Raichoor for the angular selection function of the eBOSS samples. The authors are also grateful to Pravabati Chingangbam for her carefully reading the manuscript and giving helpful comments, and to Bharat Ratra and Kin W. Ng for their stimulating comments on our results.
FD and CP are supported by the KIAS Individual Grants PG079002 and PG016904 at the Korea Institute for Advanced Study (KIAS). 
SEH is supported by the project \begin{CJK}{UTF8}{mj}우주거대구조를 이용한 암흑우주 연구\end{CJK} (``Understanding Dark Universe Using Large Scale Structure of the Universe''), funded by the Ministry of Science.
JK was supported by a KIAS Individual Grant (KG039603) via the Center for Advanced Computation at Korea Institute for Advanced Study. HSH is supported by the National Research Foundation of Korea (NRF) grant funded by the Korea government (MSIT) (No. 2021R1A2C1094577).
SA is supported by an appointment to the JRG Program at the APCTP through the Science and Technology Promotion Fund and Lottery Fund of the Korean Government, and were also supported by the Korean Local Governments in Gyeongsangbuk-do Province and Pohang City. 
The authors acknowledge the Korea Institute for Advanced Study for providing computing resources (KIAS Center for Advanced Computation) for this work.

\typeout{}
\bibliography{ap}
\appendix 

\section{Redshift-space correlation function as a standard shape}
\label{sec:beta}
\begin{figure*}[!htb]
    \centering
     \subfigure{
     \includegraphics[width=0.5\linewidth, clip]{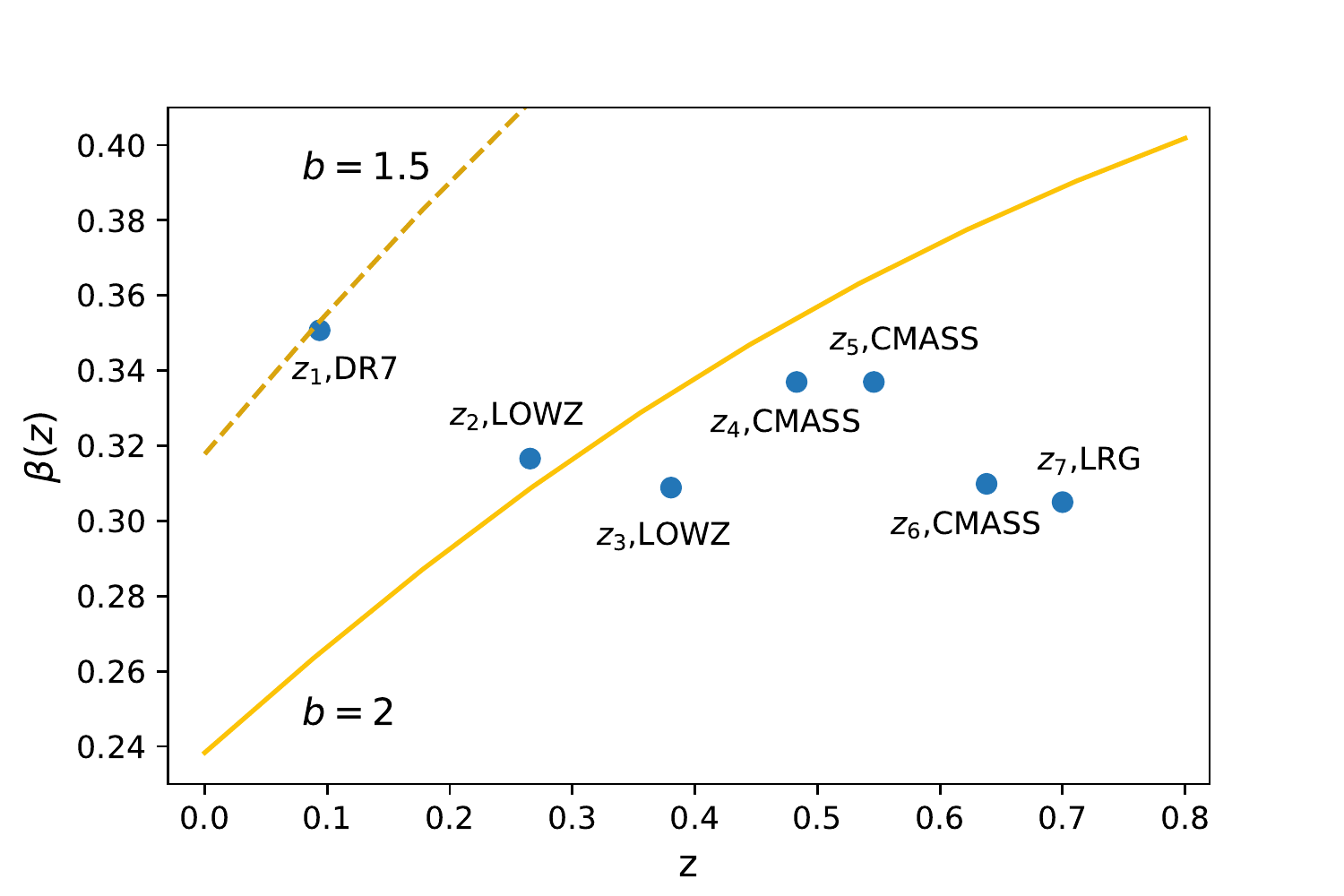}}
   \caption{The parameter $\beta(z)=f/b$ for the galaxies in the seven SDSS samples used in this study (blue dots). Note that $\beta$ is almost independent of redshift. The HR4 cosmology is used to calculate the growth rate function $f$, and the snapshots of HR4 are used to measure the matter fluctuation at various redshifts and to estimate the galaxy bias factors.  For comparison, the function $\beta(z)$ for the galaxies with a fixed bias factor of $b=1.5$ (dashed line) or 2.0 (solid line) is plotted for the HR4 cosmology.}
    \label{fig:bg}
\end{figure*}

 Let us explain why it is beneficial to use the redshift-space correlation function as a standard shape. Suppose we are given a set of galaxy redshift survey data at several redshift intervals. Let the observed galaxies have the linear bias factor that depends on redshift as given by $b(z) = (1+z)^{\alpha}$. The RSD effect due to the Kaiser effect on the shape of CF on large scales depends on the parameter $\beta(z) = f(z)/b(z).$ The growth rate function $f(z)$ depends on redshift approximately as $f(z) \propto (1+z)^{\zeta}$, where $\zeta \approx 0.82$ between $z=0$ and 0.7 in a flat $\Lambda$CDM universe with $\Omega_m =0.3$, for example. In general, one makes flux-limited surveys selecting galaxies brighter than a certain magnitude at a given redshift. And then the bias factor increases as redshift increases because brighter galaxies are selected and the clustering of underlying matter distribution becomes weaker more quickly at higher redshifts compared to that of galaxies. Therefore, the redshift dependence of the bias factor and growth rate function tend to cancel each other. In the case of the flat $\Lambda$CDM universe $\beta$ will be nearly independent of redshift if $\alpha \approx 0.82$. Then, even if the RSD effect due to peculiar velocities can be very large and distorts the shape of the CF greatly in redshift space, the shape will not evolve with redshift. In fact, as can be seen in Figure 9 the $\beta$ parameter of the SDSS galaxies used in this paper is almost independent of redshift. When the redshift-space CF is used as the standard shape, one needs to model only the weak redshift evolution of the RSD effect instead of modeling the whole RSD effect itself. Since the corrections for the nonlinear evolution and cosmology dependence are much smaller than the RSD effect itself, the redshift-space CF is a more useful standard shape than the real-space CF.

\section{angular selection function of the LRG LSS sample}
\label{sec:lrg-angular}
\begin{figure*}[!htb]
    \centering
     \subfigure{
     \includegraphics[width=1\linewidth, clip]{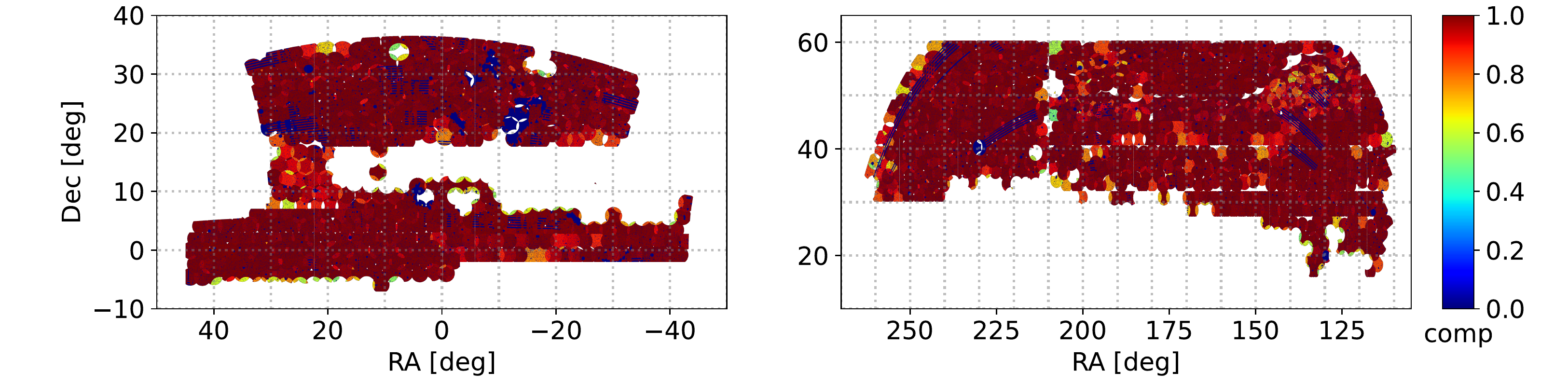}}
   \caption{Angular selection functions for the LRG clutering$\_$data samples in the north and south Galactic maps. The color bar on the right encodes the completeness of the LRG sample as a function of position on the sky.}
    \label{fig:lrg_mask}
\end{figure*}

In this section we explain how the angular selection function (veto masks $+$ tiling completeness COMP$\_$BOSS) of the LRG LSS sample is reproduced for generating mock samples.
A series of veto masks are applied to the targets of the LRG ``full" sample \citep{2020MNRAS.498.2354R}, such as LRG Collision Priority, Bad Field, and so on. However, some sectors are further removed with a completeness cut and redshift success rate cut so that there is no spurious clustering signal that is associated with plate-to-plate variations in exposure depth. 
COMP$\_$BOSS is determined on a per-sector basis as the ratio of the number of resolved fibers to the number of target in each sector, a sector being a region defined by a unique set of overlapping plates.
For recovering the completeness value of each sector, we make use of the targets in the "full" sample as there is at least one galaxy per sector. This is done with the geometry file \footnote{eBOSS$\_$QSOandLRG$\_$fulltotprintgeometry$\_$noveto.ply} which allows a mapping between polygons and sectors. Then we make use of the LSS random catalogs to decide which sector to be further removed. Finally, the angular selection functions of the NGC and SGC LRG samples used in this paper are shown in Fig.\ref{fig:lrg_mask}.

\section{Stellar Mass of eBOSS LRGs}
\label{sec:lrg-ms}
We take advantage of the availability of near-infrared photometry in the DECaLS catalogs to estimate the stellar mass of the LRGs in bands of 3.4 $\mu$m and 4.6 $\mu$m (W1, W2).
We build the spectral energy distribution of the LRGs from the eBOSS/grz-bands and W1W2-bands in Wide-field Infrared Survey Explorer (WISE, \cite{2010AJ....140.1868W}), and fit it with the CIGALE code \citep{2019A&A...622A.103B}, assuming the Chabrier \citep{2003PASP..115..763C} initial mass function, the \cite{2003MNRAS.344.1000B} stellar population synthesis model and the dust attenuation law of \cite{2000ApJ...533..682C}.
The dust emission is modelled by the simple DALE2014 module using the dust templates of \cite{2014ApJ...784...83D}.
We convert the SDSS asinh magnitudes of galaxies to fluxes in the AB system \citep{1983ApJ...266..713O} in nanomaggies as the input data of CIGALE. We also take into account of the Galactic extinction \citep{1998ApJ...500..525S}. At last, we increase the measured stellar mass  by 0.15 dex to match the stellar mass of G18, by taking into account the difference in their dust attenuation laws.
To validate our stellar mass estimate, we use the same modules to measure the stellar mass of ELG, and get consistent results with \cite{2017MNRAS.471.3955R}, G18.

\section{Fiber Collision Effect}
\label{sec:fiber}

The SDSS spectroscopic surveys are performed using fiber spectrographs. 
Because of the finite size of the fiber plugs, two objects that are closer than a minimum separation can not be targeted simultaneously. For DR7 survey, the fiber-collision scale is 55$''$. For the BOSS survey and eBOSS survey, the fiber-collision scales are  66$''$ and 62$''$, respectively.  In the presence of these collisions, the measurements of the small-scale clustering of galaxies are affected. The higher the redshift, the greater the impact, as the physical scale corresponding to the minimum fiber separation increases with redshift. Besides, the higher the galaxy density, the more close galaxy-pairs being affected. \cite{2020ApJ...897...17T} has shown that the fiber collision effect on the clustering of DR7 galaxies are mainly below the scale of 1 Mpc$/h$, which is much smaller than the scale range in our shape measurement. So we perform the fiber collision correction only for the BOSS and eBOSS galaxy samples. 

We model the fiber collision effect using the HR4 simulation. We generate a mock catalogue for BOSS within the redshift range of [0.1,0.8].  We find out all close pairs for each galaxy with separations less than 66$''$. The BOSS tile distribution achieved a tiling efficiency of $>93\%$ for all targets \footnote{\url{https://www.sdss.org/dr16/algorithms/boss_tiling/\#CollisionPriorityMask}}. So we mimic the missing galaxy effect by randomly selecting one galaxy and remove all its close pair galaxies until the total number of left galaxies is reduced to $93\%$. We measure the CF for both two catalogs, and obtain the difference between them. Then we adopt this difference to correct three moments $\nxi_{0,2,4}$ measured from observation for the fiber collision effect. We follow the same method to the LRG sample, where the fiber collisions between LRG-LRG pairs reduce the spectroscopic completeness by $\sim5\% $\citep{2017ApJ...848...76Z}. We find that the fiber effect mainly affects $\nxi_2$, while the impact on $\nxi_0$ is quite small ($\lesssim2\%$).

\section{Correlation between neighboring redshift shells}
\begin{figure*}[!htb]
    \centering
     \subfigure{
     \includegraphics[width=0.45\linewidth, clip]{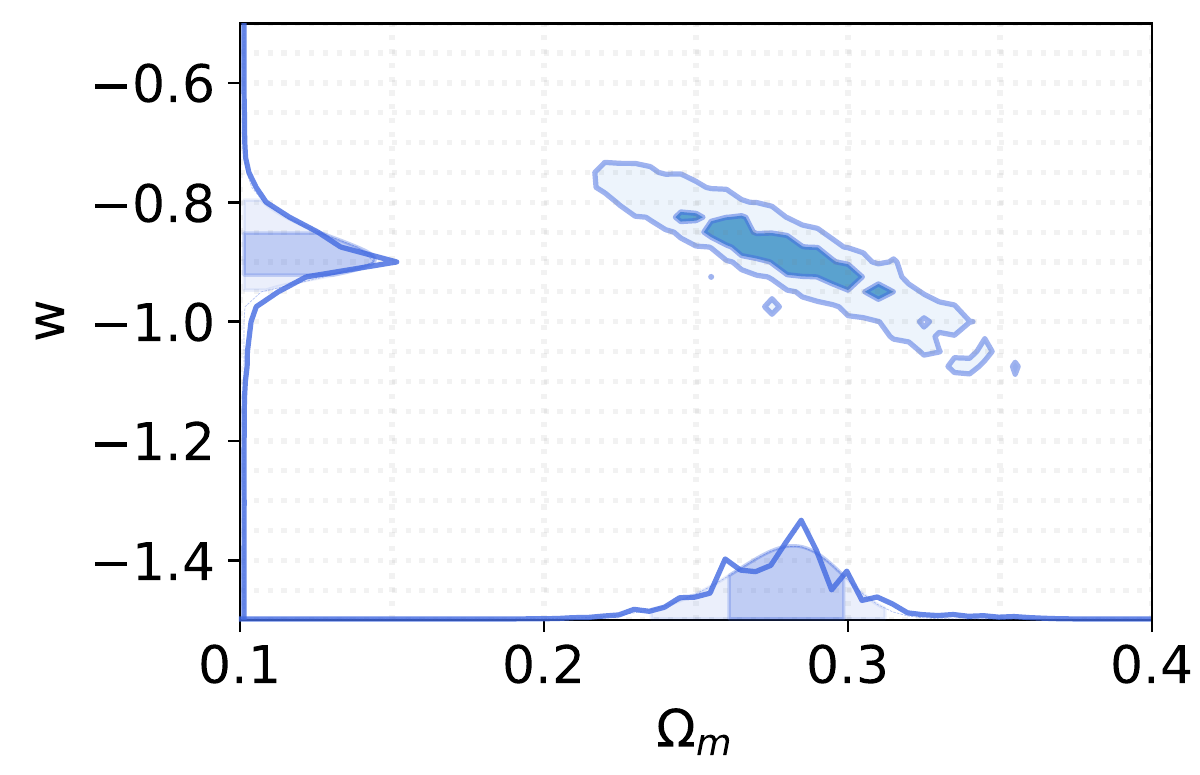}}
   \caption{Likelihood map in the $\Omega_m-w$ plane. The contours in each panel has the same meaning as the one shown in Fig. \ref{fig:result_all}, but it is derived by considering the redshift evolution of CF between all adjacent redshift slices.}
    \label{fig:neighb}
\end{figure*}

\cite{2016ApJ...832..103L,2017ApJ...844...91L,2018ApJ...856...88L} used the one-dimensional shape distortion $\Delta\nxi_{\Delta s}(z_i,z_j)$ to carry out the AP test. Here $z_i$ and $z_j$ are two different redshift bins, and $\Delta s=s_{\rm max}-s_{\rm min}$ is the range of radial integration for normalizing the CF. \cite{2019ApJ...875...92L} has improved this method by including `multi-redshifts correlation'. Each redshift bin has been in turn made the reference of the next redshift bin and a ``Full-cov" approach of likelihood was performed by jointing them all $P=(...,\Delta\nxi_{\Delta s}(z_{i-1},z_i),\Delta\nxi_{\Delta s}(z_i,z_{i+1}),...)$. This operation was performed based on two considerations that 1) all redshift evolutions $\Delta\nxi_{\Delta s}(z_{\rm ref},z_j)$ statistically correlate with each other due to the same choice of $z_{\rm ref}$; 2) the LSSs in two adjacent redshift slices are correlated with each other. The second effect must be negligibly small in the case of the AP test as the relevant scale is only about 10 $h^{-1}$Mpc while the typical sample depth is 200-500 $h^{-1}$Mpc.
In this section, we adopt similar strategy for analysis.  The result is shown in Figure \ref{fig:neighb}, which is consistent with our results presented in Figure \ref{fig:result_all}, with $\Omega_m=0.282_{-0.021}^{+0.017}$ \&. $w=-0.897_{-0.025}^{+0.045}$.
The reason is that although we have chosen one redshift as the reference in Figure \ref{fig:result_all},  the cross-correlations between different redshift bins have been taken into account in the covariance matrix of the joint vector of $P_k^{i}=(\Delta\nxi_k(z_i,z_1),...,\Delta\nxi_k(z_i,z_j),...)$ ($i\neq j$), which makes the result insensitive to the choice of reference.

\end{document}